\documentclass[a4paper,aps,prd,reprint,superscriptaddress,nofootinbib,notitlepage,showpacs,showkeys,twocolumn]{revtex4-1}

\usepackage{graphicx}
\usepackage{amsmath}
\usepackage{amssymb}
\usepackage{dcolumn}
\usepackage{bm}
\usepackage{color}
\usepackage{dsfont}
\usepackage{float}

\newcommand \beq{\begin{eqnarray}}
\newcommand \eeq{\end{eqnarray}}
\newcommand \bml{\bar M^2_{\rm L}}
\newcommand \bmt{\bar M^2_{\rm T}}

\newcommand \bgl{\bar G_{\rm L}}
\newcommand \bgt{\bar G_{\rm T}}

\renewcommand{\k}{{\bf k}}

\begin{document}
\allowdisplaybreaks

\title{The $O(N)$-model within the $\Phi$-derivable expansion to order $\lambda^2$: on the existence, UV and IR sensitivity of the solutions to self-consistent equations}

\author{Gergely Mark{\'o}}
\email{smarkovics@hotmail.com}
\affiliation{MTA-ELTE Statistical and Biological Physics Research Group, H-1117 Budapest, Hungary.}

\author{Urko Reinosa}
\email{reinosa@cpht.polytechnique.fr}
\affiliation{Centre de Physique Th{\'e}orique, Ecole polytechnique, CNRS, Universit{\'e} Paris-Saclay, F-91128 Palaiseau, France.}

\author{Zsolt Sz{\'e}p}
\email{szepzs@achilles.elte.hu}
\affiliation{MTA-ELTE Statistical and Biological Physics Research Group, H-1117 Budapest, Hungary.}

\begin{abstract}
We discuss various aspects of the $O(N)$-model in the vacuum and at finite temperature within the $\Phi$-derivable expansion scheme to order $\lambda^2$. In continuation of an earlier work, we look for a physical parametrization in the $\smash{N=4}$ case that allows us to accommodate the lightest mesons. Using zero-momentum curvature masses to approximate the physical masses, we find that, in the parameter range where a relatively large sigma mass is obtained, the scale of the Landau pole is lower compared to that obtained in the two-loop truncation. This jeopardizes the insensitivity of the observables to the ultraviolet regulator and could hinder the predictivity of the model. Both in the $\smash{N=1}$ and $\smash{N=4}$ cases, we also find that, when approaching the chiral limit, the (iterative) solution to the $\Phi$-derivable equations is lost in an interval around the would-be transition temperature. In particular, it is not possible to conclude at this order of truncation on the order of the transition in the chiral limit. Because the same issue could be present in other approaches, we investigate it thoroughly by considering a localized version of the $\Phi$-derivable equations, whose solution displays the same qualitative features, but allows for a more analytical understanding of the problem. In particular, our analysis reveals the existence of unphysical branches of solutions which can coalesce with the physical one at some temperatures, with the effect of opening up a gap in the admissible values for the condensate. Depending on its rate of growth with the temperature, this gap can eventually engulf the physical solution.
\end{abstract}


\maketitle

\section{Introduction}

The $O(N)$-model with a symmetry-breaking source $h$ is often considered in the context of light-meson phenomenology \cite{Rajagopal:1992qz, Chiku:1998kd, Patkos:2002vr, Roder:2005vt, Chivukula:1991iq}. In \cite{Marko:2013lxa}, we have studied this model at finite temperature using the so-called two-loop $\Phi$-derivable approximation. We have shown that, in a certain range of parameters, it undergoes a second order phase transition in the chiral limit ($\smash{h\to 0}$), in agreement with general expectations and in contrast to what is obtained in the Hartree approximation, where the transition is found to be first order \cite{AmelinoCamelia:1992nc, Petropoulos:1998gt, Lenaghan:1999si, Nemoto:1999qf,Bordag:2001jf, Baacke:2002pi, Reinosa:2011ut}. Within the same approximation, we have also studied to which extent the model can accommodate physical light-mesons for $\smash{N=4}$ and $\smash{h\neq 0}.$ With a parametrization based on zero-momentum curvature masses, we have found parameter values yielding physical values for the pion mass and pion decay constant and a sigma meson mass of the order of $450\,{\rm MeV}$, in line with recent dispersive analysis of $\pi\pi$ scattering data \cite{Pelaez:2015qba}. However, the narrowness of the corresponding region in parameter space and its proximity to the region where the Landau pole of the model approaches the physical scales \cite{Marko:2013lxa} can cast a doubt on the validity of our results. It is thus an interesting question to test whether corrections beyond the two-loop truncation can relax this tension between the value of the sigma meson mass and the regime of validity of the model. In this work we consider this issue within the $\Phi$-derivable expansion scheme of the two-particle irreducible (2PI) formalism to order $\lambda^2$, which adds the basketball diagram to the two-loop $\Phi$-functional of our previous work \cite{Marko:2013lxa}.

Because the critical exponents obtained in the two-loop truncation are of mean-field type, another interesting topic concerns the type of resummation needed close to the critical temperature in order to capture correctly the infrared physics determining their value. A promising resummation scheme is the 2PI-$1/N$ expansion (see e.g. \cite{Berges:2001fi}) which, beyond the leading order, contains vertex resummation at the level of the self-energy and leads to an internal propagator which exhibits a non-trivial anomalous dimension $\eta$ at criticality (for an evaluation of the exponent $\nu$ in the same approximation, see \cite{Saito:2011xq}). The value of $\eta$ calculated in \cite{Alford:2004jj} at next-to-leading (NLO) in the 2PI-$1/N$ expansion is however about $35\%-40\%$ larger than the value given by nonperturbative renormalization group methods and lattice simulations \cite{Pelissetto:2000ek}, which illustrates the difficulty of finding an adequate truncation within the 2PI formalism.

A third reason for going beyond the two-loop truncation is that the latter suffers from artifacts typical of $\Phi$-derivable approximations. In particular, although the transverse zero-momentum mass $\hat M^2_{\rm T}$ of the so-called external propagator is related to the condensate $\bar\phi$ by the symmetry constraint $\bar\phi\hat M^2_{\rm T}=h$,\footnote{In the case of a broken phase in the limit $h\to 0$, this constraint is nothing but the Goldstone theorem $\hat M^2_{\rm T}=0$.} this is not the case for the transverse zero-momentum mass $\bar M^2_{\rm T}$ of the internal propagator which appears in all the diagrams that are resummed in the approximation. There are various routes that one can follow in order to cure, or at least reduce this artifact. One possibility is to depart from the strict framework of $\Phi$-derivable approximations and to adopt the recently proposed symmetry improving procedure of Pilaftsis and Teresi, where the relation $\bar\phi\bar M^2_{\rm T}=h$ is over-imposed on the $\Phi$-derivable equations. This has been used in \cite{Pilaftsis:2013xna} in the chiral limit, both in the Hartree approximation at finite temperature and in the two-loop truncation at zero temperature. For a generalization of the symmetry improving procedure to the three-particle irreducible (3PI) framework see \cite{Brown:2015xma}. 

Although improving the symmetry content of a given truncation is a very interesting approach, here we follow a more conservative one by remaining strictly at the level of $\Phi$-derivable approximations, with the idea that the relation $\bar\phi\bar M^2_{\rm T}=h$ could be approximately recovered by increasing the order of the truncation, and we test to which extent this is true in the ${\cal O}(\lambda^2)$ $\Phi$-derivable approximation. We will see that, for not too small values of $h$, the symmetry constraint is indeed obeyed to a relatively good accuracy, but as we decrease $h$ our results depart more and more from this constraint. Interestingly however, this corresponds to a regime where the system of equations in the ${\cal O}(\lambda^2)$ truncation starts losing its (iterative) solution(s). The loss of solutions is somewhat correlated to the fact that some of the infrared modes become light (without never becoming massless) in some range of temperatures as we decrease $h$.  In particular this prohibits the access to the would-be critical region, in the chiral limit. 

In order to investigate thoroughly the loss of solution in the ${\cal O}(\lambda^2)$ truncation, we use a further approximation of the equations, that we call \emph{localization}, which is helpful in the presence of light (but not massless) modes. This type of approximation has been used previously in the literature \cite{Bordag:2000tb,Reinosa:2011cs}. As a generic feature revealed by our analysis, we find the existence of a second branch of solutions to the localized gap equation, in addition to the physical branch. This branch can coalesce with the physical one at a certain temperature, opening up a gap in the admissible values for the condensate and leading to the loss of solution to the system of  gap and field equations in the case the actual value of the condensate enters this region. It is very plausible that these considerations apply as well to the full (non-local) 2PI case, even though the iterative method that we use to solve the equations in this case does not allow us to access the other branches.\footnote{A different method, such as the Newton-Raphson algorithm, could probably allow to access unphysical branches, if they exist, provided the initial conditions are appropriately chosen.} Also, the generality of these considerations makes them interesting in view of higher orders of approximation and even other approaches dealing with non-linear integral equations, such as those encountered in the Dyson-Schwinger framework. We mention as well that our results may have an impact on the applicability of the symmetry improved procedure of Pilaftsis and Teresi refered to above. In fact, for a general truncation, and in particular for the truncations that we consider in this work, it might not be so easy to overimpose a vanishing transverse gap mass in the chiral limit without running into the difficulties that we have described in this work. In other words, the possibility of overimposing Goldstone theorem is intimately related to the ability of the truncation to cope with the IR sensitivity that this procedure introduces in the equations.

The paper is organized as follows. In Sec.~\ref{sec:main}, we gather generalities concerning the ${\cal O}(\lambda^2)$ $\Phi$-derivable approximation. In particular, we derive the corresponding equations and briefly discuss their renormalization, the details being relegated to Appendix~\ref{app:Ol2ren}. In Sec.~\ref{sec:N4}, we discuss the physical application of the model in the $N=4$ case to light-meson phenomenology by performing a parametrization based on the zero-momentum curvature masses. It turns out that, for parameters such that a sigma mass of order 400~MeV is obtained,  the scale of the Landau pole present in the model is lower than the one obtained at two-loop level. This hinders the predictive power of the model and could compromise its applicability in the context of light-meson phenomenology. We also study to which extent the symmetry constraint is obeyed by the gap mass at this level of approximation both in the range of physical values of $h$ and when approaching the chiral limit where the loss of solutions is observed. In Sec.~\ref{sec:IR-sensitivity}, using the localized approximation, we discuss the origin of the loss of solutions and compare the solutions obtained in full and localized cases. We end our paper with some final remarks, gathered in Sec.~\ref{sec:remarks} and a conclusion (Sec.~\ref{sec:conclusions}). We argue in particular that the symmetry improvement of Pilaftsis and Teresi could fail in the two-loop truncation. Appendix \ref{app:loc} presents in details the renormalization of localized equations and Appendix \ref{app:SSMmm} gives some details on the evaluation of the setting-sun diagram with two masses which appears in our investigation.

\section{The ${\cal O}(\lambda^2)$ truncation}\label{sec:main}

\subsection{Generalities}
We discuss the Euclidean $O(N)$-symmetric scalar model defined by the Lagrangian density
\beq
{\cal L}&=&\frac{1}{2}\partial_\tau\varphi_a\partial_\tau\varphi_a+\frac{\alpha}{2}\nabla\varphi_a\cdot\nabla\varphi_a\nonumber\\
&&+\frac{1}{2} m_{\rm b}^2\varphi_a\varphi_a+\frac{1}{24}\lambda_{abcd}\varphi_a\varphi_b\varphi_c\varphi_d\,,
\label{eq:Lagrangian}
\eeq
where
\beq
\lambda_{abcd}\equiv\frac{\lambda_{\rm b}}{3N}\big(\delta_{ab}\delta_{cd}+\delta_{ac}\delta_{bd}+\delta_{ad}\delta_{bc}\big)\,,
\label{Eq:bare_cpl_tensor}
\eeq
is the symmetrized coupling tensor, and $m_{\rm b}^2$ and $\lambda_{\rm b}$ denote the bare mass and the bare coupling, respectively. A summation over repeated Latin indices ($a=1,\dots,N$) is implied. The reason for introducing $\alpha$ in \eqref{eq:Lagrangian} is that we use a regularization that breaks the $O(4)$-symmetry of the Euclidean space-time at $T=0$, see \cite{Marko:2012wc,Marko:2013lxa}. It follows that the operators $(\partial_\tau\varphi_a)^2$ and $(\nabla\varphi_a)^2$ require \emph{a priori} independent renormalizations which are adjusted so that to recover the $O(4)$-symmetry at $T=0$, in the renormalized theory at large cutoff values \cite{Borsanyi:2008ar}.

In momentum space, the inverse tree-level propagator corresponding to \eqref{eq:Lagrangian} is 
\beq\label{eq:free1}
\big(G_0^{-1}\big)_{ab}(Q)=(\omega^2_n+\alpha q^2+m_{\rm b}^2)\delta_{ab}\,,
\eeq 
with $Q\equiv(\omega,{\vec q})$, $q\equiv|\vec{q}|$ and where $\omega_n\equiv2\pi n T$ denotes bosonic Matsubara frequencies defined for $n\in \mathds{Z}.$ After restricting our description to homogeneous field configurations, the 2PI effective potential (the 2PI effective action for homogeneous field and translation invariant propagators, scaled by the $4$d-volume $V/T$) can be written in the general form
\beq
\gamma[\phi,G]&=&\frac{1}{2}\int_Q^T\textnormal{tr}\big[\ln G^{-1}(Q)+G_0^{-1}(Q)G(Q)-1\big]\nonumber\\
&&\hspace{0.5cm}+\,\frac{m_{\rm b}^2}{2}\phi_a\phi_a+\gamma_\textnormal{int}[\phi,G]-h_a\phi_a\,,
\eeq
where the 2PI skeleton diagrams included in $\gamma_{\rm int}[\phi,G]$ define the truncation used to solve the model and we have introduced an explicit symmetry breaking term $h_a\phi_a$ for later use. Here, for the sum-integral we introduced the notation
\beq
\int_Q^T f(Q)\equiv T\sum_{n=-\infty}^{\infty}\int\frac{d^3q}{(2\pi)^3} f(i\omega_n,q)\,.
\eeq
Rescaling the field and the propagator as $\phi\to Z_\tau^{1/2}\phi$ and $G\to Z_\tau G$, we redefine the bare mass and the bare coupling as $m_0^2=Z_\tau m_{\rm b}^2$ and $\lambda_0=Z_\tau^2\lambda_{\rm b}$ and redefine the  inverse tree-level propagator as
\beq\label{eq:free2}
\big(G_0^{-1}\big)_{ab}(Q)\to\big(G_0^{-1}\big)_{ab}(Q)=(Z_\tau\omega_n^2+Z_s q^2+m_0^2)\delta_{ab}\,,\nonumber\\
\eeq 
where we introduced a second wave-function renormalization factor $Z_s\equiv\alpha Z_\tau,$ which reflects the explicit breaking of the $O(4)$-symmetry of the Lagrangian density \eqref{eq:Lagrangian}. Dropping an irrelevant constant arising after these redefinitions, we obtain the same expression for the 2PI effective potential as above, with $m_{\rm b}$ replaced by $m_0$ and the free propagator (\ref{eq:free1}) replaced by (\ref{eq:free2}).\\

\subsection{Gap and field equations}
The highest level of truncation that we discuss in this work contains all skeleton diagrams in the interaction term of the 2PI effective potential up to and including order $\lambda^2$. As explained in \cite{Berges:2005hc, Marko:2013lxa}, \emph{a priori} five different bare couplings $\lambda_0^{(A)},$ $\lambda_0^{(B)},$ $\lambda_2^{(A)},$ $\lambda_2^{(B)},$ and $\lambda_4$ are necessary, instead of a single one $\lambda_0,$ and a second mass parameter $m_2^2$ is also needed besides $m_0^2.$ All the bare parameters are determined by our renormalization procedure, see Appendix~\ref{app:Ol2ren}, in terms of two renormalized parameters $m_\star^2$ and $\lambda_\star.$ This truncation results in an interacting contribution to the 2PI effective potential of the form
\begin{widetext}
\beq 
\gamma_\textnormal{int}[\phi,G] &=& \frac{\hat\lambda_{abcd}}{24}\phi_a\phi_b\phi_c\phi_d+\frac{\lambda_{ab,cd}}{4}\phi_a\phi_b\int_Q^T G_{cd}(Q)+\frac{\bar \lambda_{ab,cd}}{8}\int_Q^T G_{ab}(Q)\int_P^{\rm T} G_{cd}(P) \nonumber\\ 
&-&\frac{1}{12}\int_Q^T\int_K^T\phi_a\lambda_{abcd} G_{bb'}(Q) G_{cc'}(K) G_{dd'}(-K-Q)\lambda_{a'b'c'd'}\phi_{a'}\nonumber\\
&-&\frac{1}{48}\int_Q^T\int_K^T\int_P^{\rm T}\lambda_{abcd} G_{aa'}(K)G_{bb'}(P) G_{cc'}(Q) G_{dd'}(-K-P-Q)\lambda_{a'b'c'd'}\,,
\label{Eq:gamma_int}
\eeq
\end{widetext}
given in terms of the partially symmetric coupling tensors
\begin{subequations}
\beq
\bar\lambda_{ab,cd}&\equiv&\frac{\lambda_0^{(A)}}{3N}\delta_{ab}\delta_{cd}+\frac{\lambda_0^{(B)}}{3N}\big(\delta_{ac}\delta_{bd}+\delta_{ad}\delta_{bc}\big)\,,\\
\lambda_{ab,cd}&\equiv&\frac{\lambda_2^{(A)}}{3N}\delta_{ab}\delta_{cd}+\frac{\lambda_2^{(B)}}{3N}\big(\delta_{ac}\delta_{bd}+\delta_{ad}\delta_{bc}\big)\,
\eeq
and the completely symmetric coupling tensors 
\beq
\lambda_{abcd}\equiv\frac{\lambda_\star}{3N}\big(\delta_{ab}\delta_{cd}+\delta_{ac}\delta_{bd}+\delta_{ad}\delta_{bc}\big)\,,\\
\hat\lambda_{abcd}\equiv\frac{\lambda_4}{3N}\big(\delta_{ab}\delta_{cd}+\delta_{ac}\delta_{bd}+\delta_{ad}\delta_{bc}\big)\,.
\eeq
\end{subequations}
In what follows, it is enough to restrict to propagators\\ \eject \noindent of the form
\beq
G_{ab}(Q)\equiv G_{ba}(Q)=G_{\rm L}(Q)P^{\rm L}_{ab}+G_{\rm T}(Q)P^{\rm T}_{ab}\,,
\label{eq:prop_proj}
\eeq 
where
\vglue0.75mm
\beq
P^{\rm L}_{ab}=\frac{\phi_a\phi_b}{\phi^2}\quad \textnormal{and}\quad P^{\rm T}_{ab}=\delta_{ab}-\frac{\phi_a\phi_b}{\phi^2}
\eeq
are the longitudinal and transverse projectors. We can also assume that $\phi$ is collinear to $h$ and that $h$ points along the first direction (in what follows, we use $h$ to denote the corresponding component). After the tensor structure is worked out, the potential is rewritten in terms of the components $G_{\rm L}$ and $G_{\rm T}$ as
\vglue0.5mm
\begin{widetext}
\beq
\label{eq:2loop_effpot}
\gamma[\phi,G_{\rm L},G_{\rm T}] & = & \frac{1}{2}m_2^2\phi^2+\frac{\lambda_4}{24N}\phi^4-h\phi
+\sum_{i={\rm T},{\rm L}}\frac{c_i}{2}\int_Q^T \big[\ln G^{-1}_i(Q)+(Q^2+m^2_0)\,G_i(Q)-1\big]\nonumber\\
&& +  \frac{\lambda_0^{(A+2B)}}{24N}\,{\cal T}^2[G_{\rm L}]+\frac{\lambda_0^{((N-1)A)}}{12N}{\cal T}[G_{\rm L}]{\cal T}[G_{\rm T}]+\frac{\lambda_0^{((N-1)^2 A+2(N-1)B)}}{24N}\,{\cal T}^2[G_{\rm T}]\nonumber\\
&& 
+\frac{\phi^2}{12N}\big[\lambda_2^{(A+2B)}{\cal T}[G_{\rm L}]+\lambda_2^{((N-1)A)}{\cal T}[G_{\rm T}]\big]
-\frac{\lambda^2_\star\phi^2}{36 N^2}\big[3{\cal S}[G_{\rm L}]+(N-1){\cal S}[G_{\rm L};G_{\rm T};G_{\rm T}]\big]\nonumber\\
&& - \frac{\lambda_\star^2}{144 N^2}\big[3{\cal E}[G_{\rm L}]+(N^2-1){\cal E}[G_{\rm T}]+2(N-1){\cal E}[G_{\rm L};G_{\rm T}]\big]\,,
\eeq
\end{widetext}
where $c_{\rm L} = 1$ and $c_{\rm T}=N-1.$ We also introduced the notation
\beq
\lambda_{0,2}^{(\alpha A+\beta B)}\equiv \alpha\lambda_{0,2}^{(A)}+\beta\lambda_{0,2}^{(B)}\,,
\eeq
and defined the tadpole, bubble, setting-sun, and basketball sum-integrals as:
\begin{subequations}
\beq
&&{\cal T}[G] \equiv \int_Q^T G(Q)\,,\\
&&{\cal B}[G_1;G_2](K) \equiv \int_Q^T G_1(Q)G_2(Q+K)\,,\\
&&{\cal S}[G_1;G_2;G_3](P) \equiv \int_Q^T G_1(Q){\cal B}[G_2;G_3](Q+P)\,,\\
&&{\cal E}[G_1;G_2] \equiv \int_Q^T \int_K^T G_1(Q) G_1(K) {\cal B}[G_2](K+Q)\,.\ \ \ \ \ 
\eeq
\end{subequations}
In order to alleviate the notation we used the convention that when all propagator arguments of an integral are the same, then only one argument is written. We shall also omit writing the momentum-dependence whenever the integral is taken at vanishing external momentum.\\

The usual 1PI effective potential $\gamma(\phi)$ is recovered by evaluating the 2PI effective potential as $\gamma[\phi,\bar G_{\rm L},\bar G_{\rm T}],$ that is at the solution $\bar G_{\rm L}$ and $\bar G_{\rm T}$ of the two stationarity conditions $\delta \gamma[\phi,G_{\rm L},G_{\rm T}]/\delta G_{\rm L,T}\big|_{\phi,\bar G_{\rm L},\bar G_{\rm T}}=0,$ which we call {\it gap equations}. The extrema of the effective potential obey another stationarity condition, the {\it field equation} $\delta \gamma[\phi,G_{\rm L},G_{\rm T}]/\delta \phi\big|_{\bar \phi,\bar G_{\rm L},\bar G_{\rm T}}=0.$ With the notations introduced above, the gap equations can be written as
\beq
\bar G^{-1}_{\rm L/T}(K)=K^2+\bar M^2_{\rm L/T}(K)\,,
\label{eq:gapeqsDef}
\eeq
with the momentum-dependent gap masses $\bar M^2_{\rm L/T}(K)$ given by
\begin{subequations}
\beq
\bml(K) & = & \delta Z_\tau \omega_m^2 +\delta Z_s k^2 + m^2_0+\frac{\lambda_2^{(A+2B)}}{6N}\phi^2\nonumber\\
&+&\frac{\lambda^{(A+2B)}_0}{6N}{\cal T}[\bgl]+\frac{\lambda^{((N-1)A)}_0}{6N}{\cal T}[\bgt]\nonumber\\
&-&\frac{\lambda_\star^2}{18N^2}\Big[\phi^2\big(9{\cal B}[\bgl](K)+(N-1){\cal B}[\bgt](K)\big)\nonumber\\
&+&3{\cal S}[\bgl](K)+(N-1){\cal S}[\bgl;\bgt;\bgt](K)\Big]\,\nonumber\\ \label{eq:bml}\\
\bmt(K) & = & \delta Z_\tau \omega_m^2 +\delta Z_s k^2 + m^2_0+\frac{\lambda^{(A)}_2}{6N}\phi^2+\frac{\lambda_0^{(A)}}{6N}{\cal T}[\bgl]\nonumber\\
&+&\frac{\lambda^{((N-1)A+2B)}_0}{6N}{\cal T}[\bgt]-\frac{\lambda_\star^2}{18N^2}\Big[2\phi^2{\cal B}[\bgl;\bgt](K)\nonumber\\
&+&{\cal S}[\bgt;\bgl;\bgl](K)+(N+1){\cal S}[\bgt](K)\Big]\,,\nonumber\\ \label{eq:bmt}
\eeq
and the field equation reads
\beq
h & = & \bar \phi \left[m^2_2+\frac{\lambda_4}{6N}\bar\phi^2+\frac{\lambda_2^{(A+2B)}}{6N}{\cal T}[\bgl]+\frac{\lambda_2^{((N-1)A)}}{6N}{\cal T}[\bgt]\right.\nonumber\\
&-&\left.\frac{\lambda^2_\star}{18N^2}\Big(3{\cal S}[\bgl]+(N-1){\cal S}[\bgl;\bgt;\bgt]\Big)\right].\label{eq:bphi}
\eeq
\end{subequations}

We shall also consider the curvature (zero-momentum) masses $\hat M^2_{\rm L}$ and $\hat M^2_{\rm T}$ defined from the relation
\beq
\frac{\delta^2\gamma(\phi)}{\delta\phi_a\delta\phi_b}=\hat M^2_{\rm L}P^{\rm L}_{ab}+\hat M^2_{\rm T}P^{\rm T}_{ab}\,.
\eeq
In the exact theory, these masses coincide with the gap (zero-momentum) masses, but this is generally not true in a given $\Phi$-derivable approximation. In the ${\cal O}(\lambda^2)$ truncation considered here, which preserves the exact relation
\beq
\frac{\delta^2 \gamma_\textnormal{int}[\phi,G]}{\delta\phi_a\delta\phi_b}\Big|_{\phi=0} = 2\frac{\delta \gamma_\textnormal{int}[\phi,G]}{\delta G_{ab}(Q=0)}\Big|_{\phi=0},
\label{Eq:spec_rel_trunc}
\eeq 
the gap and curvature masses coincide at $\phi=0$ \cite{Berges:2005hc}. As we have already mentioned in the Introduction, in the broken symmetry phase $\hat M_{\rm T}^2=h/\bar\phi$.

\subsection{Renormalization}
Another consequence of (\ref{Eq:spec_rel_trunc}) is that some of the bare couplings coincide, namely $m_2^2=m_0^2$ and $\lambda_2^{(A/B)}=\lambda_0^{(A/B)},$ as shown in \cite{Berges:2005hc}. The interested reader can find in Appendix~\ref{app:Ol2ren} the determination of the mass parameter $m_0^2=m_\star^2+\delta m_0^2,$ the wave-function renormalization constants $Z_\tau=1+\delta Z_\tau$ and $Z_s=1+\delta Z_s,$ and the couplings $\lambda_0^{(A/B)}=\lambda_\star+\delta\lambda_0^{(A/B)}$ and $\lambda_4=\lambda_\star+\delta\lambda_4.$  These are obtained by imposing, at a temperature $T_\star$ where the symmetry is required to be restored ($\bar\phi=0$), a set of renormalization and consistency conditions on the self-consistent propagator $G_\star(K_\star)\equiv \bar G_{\phi=0,T_\star}(K_\star)$ and the different four-point functions reviewed there ($K_\star=(\omega_\star,\k)$ with $\omega_\star=2\pi n T_\star$ the Matsubara frequency). In order to avoid any confusion, we emphasize that, in our approach, the counterterms are temperature-independent, that is at any temperature $T$ we use the same set of counterterms determined at temperature $T_\star,$ which merely plays the role of a renormalization scale.

\section{Physical parametrization for $N=4$ and ultraviolet sensitivity}\label{sec:N4}

In Ref.~\cite{Marko:2013lxa}, using the two-loop $\Phi$-derivable approximation, we investigated to which extent the $O(4)$-model with $h\ne 0$ can accommodate physical values for the pion mass and the decay constant, as well as values of the sigma meson mass in line with recent dispersive analysis of $\pi\pi$ scattering data \cite{Pelaez:2015qba}. Even though we could find parameters that matched all the constraints, the corresponding region in parameter space was quite narrow and the Landau pole present in this effective model turned out to be relatively close to the highest physical scale of the model, that is the sigma mass. In such a situation the observables depend on the regulator (shape and cutoff scale), which hinders the predictivity of the model. In this section, we use the ${\cal O}(\lambda^2)$ $\Phi$-derivable approximation to investigate whether this tension observed within a two-loop truncation can be relaxed by higher order corrections.

\subsection{Solving the equations}
We solve the ${\cal O}(\lambda^2)$ truncation using the same numerical framework as in the two-loop truncation \cite{Marko:2013lxa,Marko:2012wc}, that is we use a $N_\tau\times N_s$ grid to store the propagator(s) evaluated at $N_\tau-1$ positive Matsubara-frequencies beyond the zero mode, and at $N_s$ non-zero $3$-momentum values, of which the largest is $\Lambda$. Throughout this paper we use the following discretization: $N_\tau=2N_s=2^{12}$ and $\Lambda/T_\star=10$ ($T_\star=1$ is used in the code). Also, if not stated otherwise, we use $m_\star^2/T_\star^2=0.04$ and $\lambda_\star=3$ in order to illustrate some features of the model. Convolutions of rotationally invariant functions of $3$-momenta and Matsubara frequencies are calculated using fast Fourier transform algorithms and integrals in momentum-independent quantities are evaluated with the extended trapezoidal rule. The algorithm needs a slight adjustment as compared to the one we used in the two-loop case, since in the ${\cal O}(\lambda^2)$ case a self-consistent equation for $G_\star,$ namely \eqref{Eq:Gstar_finite}, has to be solved already at $T_\star,$ while in the two-loop case $G_\star$ is just a free propagator.

One important ingredient in solving iteratively a self-consistent equation is the so-called under-relaxation method, which is meant to extend the domain of convergence of the iterative method and can be applied to the (inverse) propagator, as in \cite{Berges:2004hn, Marko:2012wc}, or to the self-energy, as in \cite{Baacke:2004dp, Marko:2013lxa}. Technically the method amounts to using at $i$th order of the iteration a weighted average of the $(i-1)$th and $i$th quantities, with some parameter $\alpha<1.$ In this work we apply this method to update a generic squared gap mass $\bar M^2(Q)$, which up to a constant is the self-energy, as follows
\beq
\bar M_\textnormal{update,(i)}^2(Q)=\alpha\bar M^2_{(i)}(Q)+(1-\alpha)\bar M^2_{(i-1)}(Q)\,.
\label{Eq:under-relax}
\eeq

\begin{figure}[!b]
\includegraphics[width=0.48\textwidth]{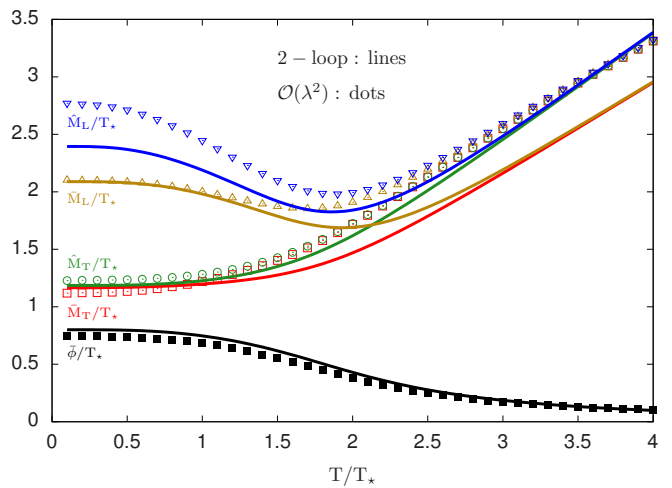}
\caption{Comparison between the solutions of the ${\cal O}(\lambda^2)$ and two-loop truncations. We show the temperature dependence of the condensates, (zero-momentum) curvature and gap masses of the longitudinal and transverse modes for $N=4$ and parameter values $m_\star^2/T_\star^2=0.252,$ $\lambda_\star=28.871,$ and $h/T_\star^3=1.125$. \label{Fig:2PI_N_BB_v_2L}}
\end{figure}


The coupled field and gap equations are solved as a function of the temperature $T$ by updating them sequentially. In the physical case ($h\ne0$) the result is shown in Fig.~\ref{Fig:2PI_N_BB_v_2L} for $N=4$ and a parameter set at which the parametrization of the model was done in the two-loop truncation based on the curvature, zero-momentum masses $\hat M_{\rm L}$ and $\hat M_{\rm T},$ as described in \cite{Marko:2013lxa}. We observe that the restoration of chiral symmetry is reflected differently at the level of the temperature evolution of the gap and curvature masses in the ${\cal O}(\lambda^2)$ approximation compared to the two-loop one. In the  ${\cal O}(\lambda^2)$ approximation the curvature and gap masses approach each other with increasing temperature in both the longitudinal and the transverse sectors and all these masses becomes practically degenerate at large values of the temperature. In contrast, in the two-loop approximation the gap masses approach each other with increasing $T$ and the same happens with the curvature masses, but at high temperature the value of the gap mass is different from the value of the curvature mass. Although in the ${\cal O}(\lambda^2)$ approximation the equality between curvature and gap masses only holds identically at $\phi=0$, in practice we observe that the gap and curvature masses of the respective transverse or longitudinal sectors approach each other before the masses of the longitudinal sector approach those of the transverse sector.

\subsection{Landau pole and the value of the sigma mass}
Based on the result shown in Fig.~\ref{Fig:2PI_N_BB_v_2L}, namely that $\hat M_{\rm L}$ is larger in the ${\cal O}(\lambda^2)$ case than in the two-loop one, one could think that in the former case the model could accommodate larger values of $\hat M_{\rm L}$ than it was possible to reach in the two-loop case, where values above $400$~MeV were obtained only in a small region of the parameter space. However, it turns out that this expectation is not fulfilled when one actually does the physical parametrization of the model in the ${\cal O}(\lambda^2)$ truncation in exactly the same way as it was done in \cite{Marko:2013lxa} in the two-loop truncation, that is requiring at $T=0$ the (zero-momentum) transverse curvature mass and condensate to have the value of the pion mass and decay constant, respectively. To ease our search for parameters satisfying these conditions we allow for $1\%$ variation from the value of the pion mass, and require $\hat M_{\rm T,0}=138\pm 1.38$~MeV, while $\bar\phi_0=93$~MeV determines the value of $T_\star$ in MeV, as the dimensionful quantities are measured in units of $T_\star$ and $T_\star$ is set to 1 in the code. 

\begin{figure}[!t]
\includegraphics[width=0.48\textwidth]{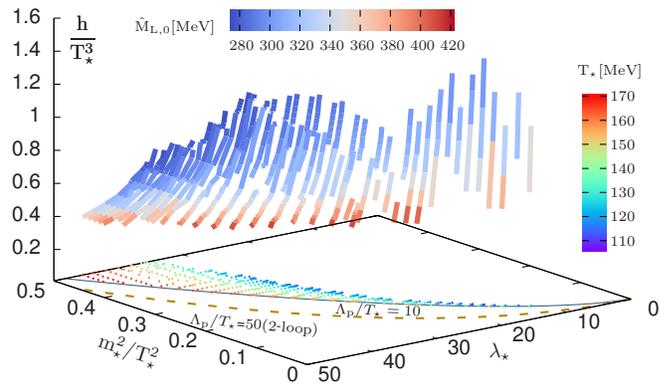}
\caption{Light-meson phenomenology based parametrization of the model for $N=4$ and $h\ne0.$ Shown are points in the parameter space which satisfy the parametrization criteria in the  ${\cal O}(\lambda^2)$ truncation (see the text). The palettes show the value of $T_\star$ and of $\hat M_{\rm L}$ at $T=0$. \label{Fig:parametrization}}
\end{figure}

The result of the parametrization is shown in Fig.~\ref{Fig:parametrization}, where only those points of the parameter space are presented for which the relation $\hat M_{\rm L,0}>2\hat M_{\rm T,0}$ is satisfied. This latter condition eliminates for $m^2_\star/T_\star^2>0.35$ points with large $h$ and small $\lambda_\star$. The vacuum values are extrapolated from values measured at $T/T_\star=0.2,$ 0.25, and 0.3. One can see that similar values of $\hat M_{\rm L,0}$ are achieved as in Fig.~2 of Ref.~\cite{Marko:2013lxa} obtained within the two-loop truncation, however, they occur in the present case at parameters for which the scale $\Lambda_{\rm p}$ of the Landau pole is lower than it was in the two-loop truncation and thus closer to the physical scales: $\Lambda_{\rm p}/T_\star=50$ in the two-loop case (dashed curve in the top panel of Fig.~\ref{Fig:parametrization}) versus $\Lambda_{\rm p}/T_\star=10$ in the present case (solid curve in the same panel), while the values of $T_\star$ are similar in the two truncation schemes ($T_\star$ can be read from the vertical palette in the top panel of Fig.~\ref{Fig:parametrization}). 

Usually the Landau pole $\Lambda_{\rm p}$ is defined as the value of the cutoff where for fixed values of parameters $m^2_\star/T^2_\star$ and $\lambda_\star$ a singularity is observed in the relation connecting the bare and renormalized parameters, as the cutoff is varied. In the two-loop case, where the relations between the bare and renormalized couplings were simpler than those in Eqs.~\eqref{Eq:l0B_at_zero_mom} and \eqref{Eq:l0A_at_zero_mom} of the  ${\cal O}(\lambda^2)$ truncation, we could show analytically that $\lambda_0^{(A)}$ is the first coupling which diverges as $\Lambda$ increases. In the present truncation we observe numerically that this is still the case. Since we work with a fixed cutoff set to $\Lambda/T_\star=10$ and in the process of parametrization a range of the renormalized $\lambda_\star$ is scanned\footnote{We mention that due to the growing of the bare couplings as $\lambda_\star$ increases, one needs to lower the parameter $\alpha$ of Eq.~\eqref{Eq:under-relax} used in the iterative method. To reach convergence, typically one needs smaller values of $\alpha$ at temperatures of the order of the pseudocritical temperature ($\alpha=0.1$) than those used at small temperatures ($\alpha=0.25\dots 0.5$).}, the scale of the Landau pole becomes the value of the cutoff used at the value of $\lambda_\star$ where a singularity is observed in the bare coupling $\lambda_0^{(A)}$.\footnote{In principle, by varying the cutoff at fixed $m_\star^2$ and $\lambda_\star,$ one could obtain the Landau pole as the value of the cutoff where the bare coupling $\lambda_0^{(A)}$ changes sign.} For $\lambda_\star$ which are bigger than this value, we observe that the longitudinal curvature mass squared $\hat M_{\rm L,0}^2$ becomes negative, which signals an instability in the model, preventing its parametrization and thus its physical applicability. The same conclusion applies for the case when at fixed $\lambda_\star$ we try push the cutoff above the value of $\Lambda_{\rm p}$ corresponding to this coupling. 

From the above discussion on the Landau pole follows that we can also look at the result presented in  Fig.~\ref{Fig:parametrization} from another point of view: values of the sigma mass above 400~MeV are obtained for cutoff values close to the Landau pole. As already mentioned, this result hinders the predictive power of the $O(4)$-model as far as light-meson phenomenology is concerned because it is difficult to obtain predictions which do not depend on the details (shape, cutoff scale) of the regulator used. We mention however that our parametrization is based on zero-momentum (curvature) masses whereas, strictly speaking, it should be based on pole masses which are the physical, renormalization invariant masses. Our approach implicitly assumes that, in the considered approximation and within our renormalization scheme, these two definitions of the mass are not far from each other. A more thorough investigation would require solving the propagator equations in Minkowski space and study the analytic structure of the propagator because the sigma particle appears in fact as a complex pole on the second Riemann sheet. The relation between the pole and curvature mass of the pion was investigated recently in \cite{Helmboldt:2014iya} within the functional renormalization group approach. We note also that, our renormalization scheme is such that the system is in the symmetric phase at the temperature $T=T_\star$. Other schemes allow one to renormalize the theory in the broken phase directly at $T=0$ \cite{Destri:2005qm, Jakovac:2008zq, Fejos:2007ec}. It would be interesting to redo the parametrization in those schemes and check whether the features observed here persist.

\subsection{Symmetry constraint}
\begin{figure}[!t]
\includegraphics[width=0.48\textwidth]{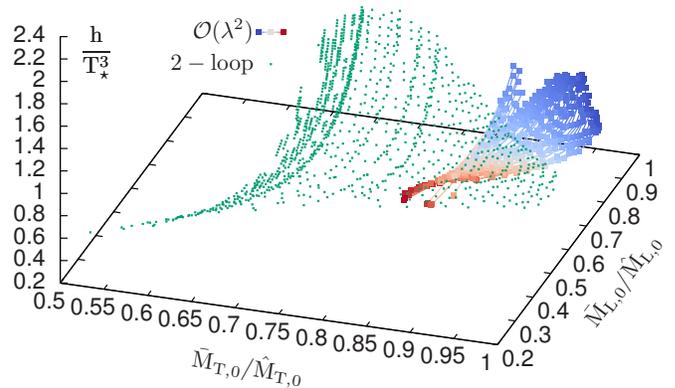}
\caption{Comparison between the ${\cal O}(\lambda^2)$ and two-loop truncations concerning the restoration of the exact identity between curvature and gap masses, based on the parameters for which the parametrization criteria are satisfied at $N=4$ and $h\ne0$ in the respective truncations. \label{Fig:constraint}}
\end{figure}

Figure~\ref{Fig:2PI_N_BB_v_2L} shows that, as opposed to the longitudinal sector, in the transverse sector the gap mass remains close to the curvature mass $\hat M_{\rm T}\equiv \big(h/\bar\phi\big)^{1/2}$ even deep in the broken phase, in both two-loop and ${\cal O}(\lambda^2)$ truncations. Actually, for that particular set of parameters, deep in the broken phase the curvature mass is closer to the gap mass in the two-loop truncation, which is not in line with the expectation that the symmetry constraint, that is the agreement between gap and curvature masses, is better satisfied with increasing order of the truncation. For this reason it is interesting to compare the ratio of the gap to curvature mass for points of the parameter space where in the respective truncation the parametrization criteria are satisfied. This is done in Fig.~\ref{Fig:constraint}, where one sees that indeed, in the  ${\cal O}(\lambda^2)$ truncation this ratio is generally closer to $1$ than in the case of the two-loop truncation in both transverse and longitudinal sectors.

\begin{figure}[!b]
\includegraphics[width=0.49\textwidth]{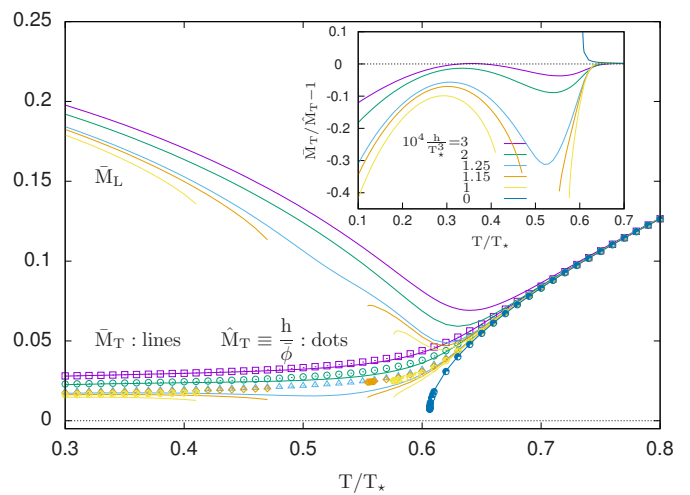}
\caption{The gap masses and the transverse curvature mass as a function of $T$ obtained in the ${\cal O}(\lambda^2)$ truncation at $N=4$ and decreasing values of the external source $h.$ No solution is found for small $h$ in some temperature range. In the chiral limit the solution is lost with decreasing $T.$ The inset shows that the relative difference between the transverse curvature and gap masses increases with decreasing $h.$ \label{Fig:full-2PI_BB_N4_h-dep}}
\end{figure}

Also, according to Fig.~\ref{Fig:2PI_N_BB_v_2L}, deep in the broken phase of both the ${\cal O}(\lambda^2)$ and two-loop truncations the transverse gap mass is smaller than the curvature mass. We know from Ref.~\cite{Marko:2013lxa} that in the chiral limit $\bar M_{\rm T}$ stays finite while $\hat M_{\rm T}$ vanishes, which means that, starting from physical parameters, $\bar M_{\rm T}$ and $\hat M_{\rm T}$ reverse order as the value of the external field $h$ is lowered. It is therefore interesting to check what happens with the gap mass as we decrease $h$. This is shown in Fig.~\ref{Fig:full-2PI_BB_N4_h-dep} where we observe that the fulfillment of the symmetry constraint becomes worst as one decreases $h$. However, this also corresponds to values of $h$ where a loss of solution is observed in some range of temperatures.

The temperature where the solution of the model is lost with decreasing (resp. increasing) temperature is higher (resp. smaller) the smaller the value of $h.$ In contrast, for large values of $h$, the solution exists down to small temperatures and $\hat M_{\rm T}\approx\bar M_{\rm T}$ even when $\bar\phi$ is not small anymore. This latter feature persists when lowering $h$, however, the difference between the transverse curvature and gap masses increases with decreasing $h$ in some temperature range.  Also, if the temperature is decreased in the chiral limit, the solution in the symmetric phase is lost at some value of temperature where the masses are still relatively large. 

Entering the broken phase in the chiral limit and for $\smash{N=4}$ proved impossible with the iterative method used, both by decreasing the temperature at $\smash{h=0}$, or by decreasing $h$ at some fixed values of the temperature. This is due to the lack of convergence of the iterative method whenever the equations become sensitive to the IR. While decreasing $h$ at fixed $T$ we had to lower the parameter $\alpha$ used in the under-relaxation method (see Eq.~\eqref{Eq:under-relax}), which increased substantially the number of iterations and the computation time, and therefore we gave up the procedure at $\alpha=10^{-3}$. Entering the broken phase was possible, however, in the chiral limit of the one-component case, where as we shall see in the next section, the broken phase solution is lost by increasing the temperature. Similarly to the $\smash{N=4}$ case, the symmetric phase solution is also lost there as the temperature is decreased.

\section{Loss of solutions\\ and infrared sensitivity}\label{sec:IR-sensitivity}
In the present ${\cal O}(\lambda^2)$ truncation, where the gap mass coincides with the curvature mass at $\smash{\phi=0}$, the loss of symmetric phase solution at a temperature where the gap mass is non-zero implies that we cannot decide about the  second order nature of the phase transition in the chiral limit along the lines of our previous investigations in Refs.~\cite{Reinosa:2011ut, Marko:2012wc, Marko:2013lxa}. This feature deserves further analysis and we will try to understand it more thoroughly, in particular its infrared nature and its relation to the existence of multiple branches of solutions to the gap equation. To this purpose, we shall approximate the gap and field equations by localizing the momentum-dependent masses. This allows for a semi-analytical understanding of the above mentioned features.

In this section we first give a natural, although to some degree heuristic, recipe to obtain finite localized equations. Next we show in the two-loop truncation, both in the $N=1$ and $N=4$ cases, that the localized equations reproduce features of the original equations and, therefore, that the localization procedure can be applied to investigate the IR sensitivity of the full 2PI equations in the ${\cal O}(\lambda^2)$ truncation. The investigation based on the localized equations reveals the existence of a gap in the values of the field for which the gap equations do not have a solution. In the two-loop approximation the minimum of the potential lies outside this forbidden range and, as a result, a solution to the coupled field and gap equations exists for all values of the temperature, while in the ${\cal O}(\lambda^2)$ truncation the minimum of the potential enters the forbidden region, which leads to the loss of solution in some temperature range. 

\subsection{Localized approximations}\label{Sec:LocDef}

Local approximations of a set of nonlocal equations have been used several times in the literature \cite{Bordag:2000tb,Reinosa:2011cs}. The idea is that in any regime where the infrared modes become light, integrals are dominated by the value of the propagator in the vicinity of $\smash{Q=0}$. A good qualitative (and sometimes even quantitative) approximation is then to replace the full momentum-dependent masses $\bar M^2_{\rm L,T}(Q)$ by local ones $\bar M^2_{\rm L,T}$ which obey the same equations at $Q=0$.\footnote{More generally one can replace the momentum dependent masses $\bar M^2_{\rm L,T}(Q)$ by their Taylor expansion around $\smash{Q=0}$, $\bar M^2_{\rm L,T}+\alpha_{\rm L,T}\omega^2_n+\beta_{\rm L,T} q^2+\dots$, to some order. The coefficients $\alpha_{\rm L,T},\,\beta_{\rm L,T},\,\dots$ are obtained by evaluating derivatives (resp. finite differences) of the gap equation for $\bar M^2_{\rm L,T}(K)$ with respect to $k^2$ (resp. $\omega^2_n$), evaluated at $K=0$. This bears some resemblance to the derivative expansion used in the context of the functional renormalization group.} For instance, localizing the $\smash{N=1}$ bare gap equation of the ${\cal O}(\lambda^2)$ truncation yields
\beq\label{eq:loc_ex}
\bar M^2 &=&m_0^2+\frac{\lambda_2}{2}\phi^2+\frac{\lambda_0}{2}{\cal T}[\bar G]-\frac{\lambda_3^2\phi^2}{2}{\cal B}[\bar G]-\frac{\lambda_4^2}{6}{\cal S}[\bar G],\nonumber\\
\eeq
where we defined $\smash{\bar G(Q)\equiv 1/(Q^2+\bar M^2)}$ and we used our convention not to write the external momentum when it is taken equal to zero. The choice of the bare couplings $\lambda_0$, $\lambda_2$, $\lambda_3$ and $\lambda_4$ in (\ref{eq:loc_ex}) will be explained below.

It is important to stress that local approximations make sense in situations where the infrared modes become light in some regime, but not massless. This is precisely the case in this work and also in \cite{Reinosa:2011cs}. In situations where the lowest modes become massless and the propagator develops an anomalous dimension, localizing the equations obviously misses this feature, see Sec.~\ref{sec:lim_loc} for more details. We mention also that another way to obtain local approximations is to consider the  so-called two-particle point-irreducible (2PPI) formalism \cite{Verschelde:2000dz}, which gives diagrams of different topology compared to those of the 2PI approximation at the level of the gap masses (namely dressed tadpole-like diagrams). An important difference with the present approach is that the 2PPI formalism is a systematic expansion scheme of the original theory whereas the localization of 2PI approximations corresponds to an approximation of a systematic expansion scheme. The latter is nevertheless useful to understand certain features of the 2PI expansion scheme, some of them analytically or with a simple numerical treatment which allows to locate all the solutions, not only those reachable iteratively. Below, we shall employ localization to understand certain features of the two-loop and ${\cal O}(\lambda^2)$ $\Phi$-derivable approximations, in particular the loss of solution that we observed in the latter case.\\

Before we proceed, we need to say a few words about the renormalizability of localized equations such as (\ref{eq:loc_ex}). The choice of bare couplings $\lambda_0$, $\lambda_2$, $\lambda_3$, $\lambda_4$, $\dots$ depends on the adopted point of view. In some specific applications, such as in \cite{Reinosa:2011cs}, the couplings can all be taken equal to a finite value $\lambda_\star$. Instead, if one is interested in obtaining a renormalized version of bare localized equations such as (\ref{eq:loc_ex}), one important question is how to choose the bare parameters in order to transform the bare localized equations into finite localized ones. In Appendix~\ref{app:loc}, we show how to renormalize the bare gap equation to all orders in the $\smash{N=1}$ case using temperature and field independent bare parameters. For instance, upon appropriate choice of $m^2_0$, $\lambda_0$, $\lambda_2$, $\lambda_3$ and $\lambda_4$, the bare localized equation (\ref{eq:loc_ex}) can be put into the renormalized form
\beq\label{eq:loc_ex_ren}
\bar M^2 &=&m_\star^2+\frac{\lambda_\star}{2}\phi^2+\frac{\lambda_\star}{2}{\cal T}_{\rm F}[\bar G]-\frac{\lambda_\star^2\phi^2}{2}{\cal B}_{\rm F}[\bar G]-\frac{\lambda_\star^2}{6}{\cal S}_{\rm F}[\bar G]\,,\nonumber\\
\eeq
where ${\cal T}_{\rm F}[\bar G]$, ${\cal B}_{\rm F}[\bar G],$ and ${\cal S}_{\rm F}[\bar G]$ are the following UV finite combinations of sum-integrals:
\beq
\label{eq:TadF}
{\cal T}_{\rm F}[\bar G]&\equiv&{\cal T}[\bar G]-{\cal T}_\star[G_\star]-(\bar M^2-m_\star^2)\frac{d{\cal T}_\star[G_\star]}{dm^2_\star}\,,\\
{\cal B}_{\rm F}[\bar G]&\equiv&{\cal B}[\bar G]-{\cal B}_\star[G_\star]\,,
\label{eq:BubF}
\eeq
and
\beq
{\cal S}_{\rm F}[\bar G] & = & {\cal S}[\bar G]-{\cal S}_\star[G_\star]-(\bar M^2-m^2_\star)\frac{d{\cal S}_\star[G_\star]}{dm_\star^2}\nonumber\\
& - & 3{\cal T}_{\rm F}[\bar G]{\cal B}_\star[G_\star]\,,
\label{eq:SSF}
\eeq
where it is understood that a sum-integral with a subscript $\star$ is evaluated at the temperature $T_\star$ with the propagator $G_\star(Q_\star) \equiv 1/(Q_\star^2+m^2_\star).$ Equation (\ref{eq:loc_ex_ren}) illustrates a generic feature of our procedure for renormalizing the localized bare gap equation in the $\smash{N=1}$ case: the renormalized equation has the same form as the bare one, except that bare parameters are replaced by renormalized ones and that any diagram ${\cal D}[\bar G]$ in the bare equation is replaced by a UV finite version ${\cal D}_{\rm F}[\bar G]$ constructed from ${\cal D}[\bar G]$ by using a systematic rule described in Appendix~\ref{app:loc} and illustrated above for the cases ${\cal D}={\cal T}$, ${\cal B}$ and ${\cal S}$. This rule generalizes the one obtained from the method developed in \cite{Fejos:2007ec} to determine the counterterms in various models and at the lower orders of the $\Phi$-derivable expansion scheme.

As we argue in Appendix~\ref{app:loc}, the procedure used to renormalize the propagator equation at $N=1$ does not work whenever we couple the gap and field equations or we consider the case $\smash{N\neq 1}$. Nevertheless, the renormalization procedure for the gap equation in the $\smash{N=1}$ case allows us to construct a natural recipe to associate to a given $\Phi$-derivable approximation a finite localized approximation. For instance, the two-loop and ${\cal O}(\lambda^2)$ truncations to be considered below,  involve the sum-integrals ${\cal B}[\bgl;\bgt]$ and ${\cal S}[\bgl;\bgt;\bgt]$ which, according to our recipe, should be replaced by
\beq
{\cal B}_{\rm F}[\bgl;\bgt]&\equiv&{\cal B}[\bgl;\bgt]-{\cal B}_\star[G_\star]
\eeq
and
\beq
{\cal S}_{\rm F}[\bgl;\bgt;\bgt]&\equiv&{\cal S}[\bgl;\bgt;\bgt]-{\cal S}_\star[G_\star]
\nonumber\\&-&
\frac{1}{3}\left[2(\bar M^2_{\rm T}-m^2_\star) + \bar M^2_{\rm L}-m^2_\star \right]\frac{d{\cal S}_\star[G_\star]}{dm^2_\star}
\nonumber\\&-& 
(2{\cal T}[\bgt]+{\cal T}[\bgl]){\cal B}_\star[G_\star]\,.\quad 
\eeq 
All integrals appearing in ${\cal T}_{\rm F}[\bar G]$, ${\cal B}_{\rm F}[\bar G]$, ${\cal S}_{\rm F}[\bar G]$, ${\cal B}_{\rm F}[\bgl;\bgt]$, and ${\cal S}_{\rm F}[\bgl;\bgt;\bgt]$ can be computed with less numerical effort than those appearing in the full 2PI equations using either dimensional or cutoff regularization. In the present study we use a cutoff regularization scheme in which the modulus of the spatial momentum of every propagator is cut off.  In particular, this allows us to maintain all our bare couplings positive and thus to avoid unphysical features related to the Landau pole. All integrals with one type of propagator can be found in Appendix~B2 of \cite{Marko:2012wc} and since for $G_{\rm L}\ne G_{\rm T}$ the bubble integral ${\cal B}[G_{\rm L};G_{\rm T}]$ is computed as a difference of tadpoles over a difference of squared masses, the only integral of the localized approximations not appearing in our previous papers within a cutoff regularization is ${\cal S}[G_{\rm L};G_{\rm T};G_{\rm T}]$. The expression of this setting-sun integral is given in Eq.~\eqref{Eq:SSMmm_3d_cut} of Appendix~\ref{app:SSMmm}.

\subsection{Full vs localized two-loop truncation}\label{Sec:Compare2}

In order to get some insight on how good the localized approximation can be in capturing qualitative (and sometimes even quantitative) features of the full equations, it is interesting to solve the corresponding equations in the two-loop truncation and compare to the results of the full two-loop truncation investigated in Refs.~\cite{Marko:2012wc,Marko:2013lxa}. This will also allow us to introduce, in a simpler context, some of the concepts that we shall be using later on and which eventually help understanding the loss of solutions observed in the ${\cal O}(\lambda^2)$ case. 

The coupled system of finite localized two-loop gap and field equations for any $N$ reads
\begin{subequations}
\label{Eq:O2l_local}
\beq
\bml&=&m_\star^2+\frac{\lambda_\star}{2N}\left(\phi^2+{\cal T}_{\rm F}[\bgl]\right)+(N-1)\frac{\lambda_\star}{6N}{\cal T}_{\rm F}[\bgt]\nonumber\\
&&-\frac{\lambda_\star^2\phi^2}{18N^2}\left(9{\cal B}_{\rm F}[\bgl]+(N-1){\cal B}_{\rm F}[\bgt]\right),\label{eq:O2lNfinGapL}\\
\bmt&=&m_\star^2+\frac{\lambda_\star}{6N}\left(\phi^2+{\cal T}_{\rm F}[\bgl]\right)+(N+1)\frac{\lambda_\star}{6N}{\cal T}_{\rm F}[\bgt]
\nonumber\\
&&-\frac{\lambda_\star^2\phi^2}{9N^2}{\cal B}_{\rm F}[\bgl;\bgt]\,,\label{eq:O2lNfinGapT}\\
h&=&\bar\phi\left[m_\star^2+\frac{\lambda_\star}{6N}\bar\phi^2+\frac{\lambda_\star}{2N}{\cal T}_{\rm F}[\bgl]+(N-1)\frac{\lambda_\star}{6N}{\cal T}_{\rm F}[\bgt]\right.\ \ \nonumber\\
&&\left. -\frac{\lambda_\star^2}{18N^2}\left(3{\cal S}_{\rm F}[\bgl]+(N-1){\cal S}_{\rm F}[\bgl;\bgt;\bgt]\right)\right].\label{eq:O2lNfinField}
\eeq
\end{subequations}

\begin{figure}[!t]
\includegraphics[width=0.48\textwidth]{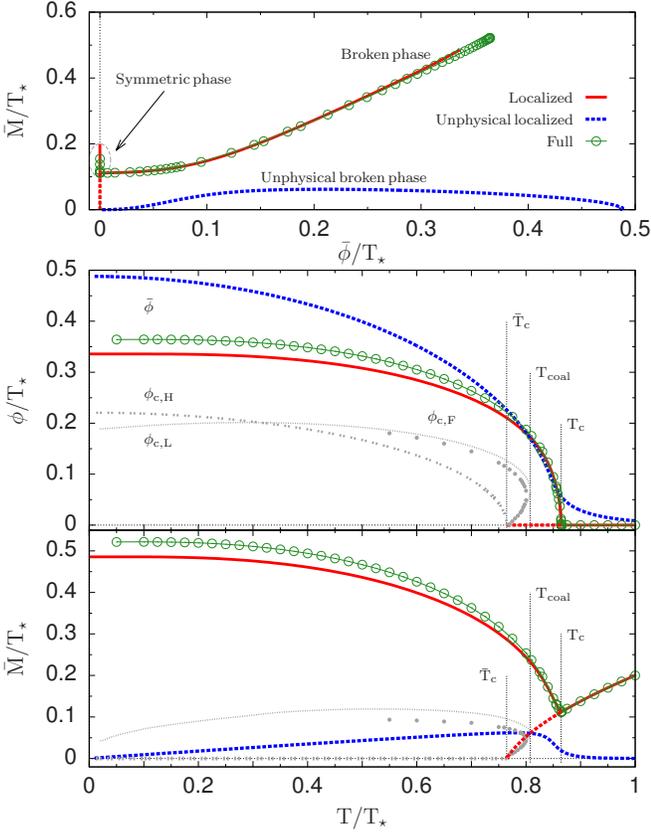}
\caption{Comparison of the full and localized results in the two-loop truncation for $N = 1$ and $h = 0$. Top panel: parametric representation $T\mapsto (\bar\phi(T),\bar M(T))$ of the solutions to the system of gap and field equations, as the temperature is varied. Middle panel: solutions of the field equation as a function of $T$, together with the $\phi_{\rm c}$ line (in units of $T_\star$) of various approximations, which limits the range of $\phi$ where the respective gap equation has a solution. The subscripts F, L, and H denote curves obtained in the full two-loop, localized two-loop and Hartree truncations, respectively. Bottom panel: the gap mass as a function of $T$.  In all panels the dashed red line corresponds to the solution of the gap equation at $\phi=0$ in the broken symmetry phase.\label{Fig:2l_local_v_full_N1}}
\end{figure}

The comparison between the localized and the full 2PI results is shown in Fig.~\ref{Fig:2l_local_v_full_N1} for the chiral limit of the $\smash{N=1}$ case. The middle and bottom panels represent the solutions $\bar M(T)$ and $\bar\phi(T)$ which are combined into a parametric plot $T\mapsto(\bar M(T),\bar\phi(T))$ in the top panel. We see that the masses agree in the symmetric phase where the two approximations coincide, while, the deeper we go into the broken phase, the bigger the difference is. We also see that, as the temperature is decreased, a nontrivial solution to the field equation appears below $T_{\rm c},$ while $\phi=0$ remains a solution only down to $\bar T_{\rm c}$. For $T<\bar T_{\rm c}$ the gap equation admits a solution only above some ``critical'' value of the field  denoted by $\phi_{\rm c}$. This was the case already in the Hartree approximation (see Refs.~\cite{Reinosa:2011ut} and \cite{Marko:2013lxa}), but interestingly, in the two-loop approximation $\phi_{\rm c}$ turns out to be bivalued for some temperature range between $\bar T_{\rm c}$ and the turning point of the curve occurring at a temperature denoted as $T_{\rm coal}$ for a reason which will become clear below.\footnote{We mention that, as opposed to the Hartree case where the gap mass vanishes along $\phi_{\rm c, H},$ both in the full and localized two-loop truncations the use of $\phi_{\rm c}$ is an abuse of notation because in fact $\bar M_{\phi_{\rm c}}$ never vanishes, as indicated in the lower panel of Fig.~\ref{Fig:2l_local_v_full_N1}.}

\begin{figure}[!t]
\includegraphics[width=0.48\textwidth]{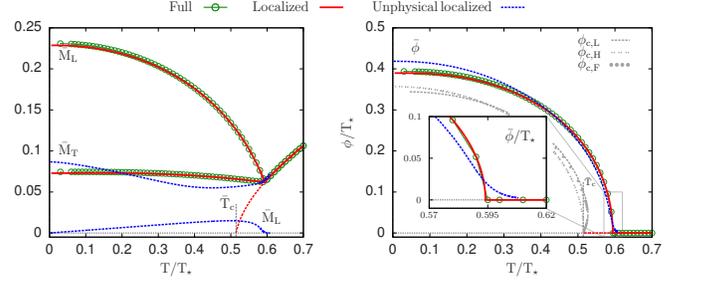}
\caption{The gap masses (left panel) and the field expectation value (right panel) as a function of the temperature in the localized and full two-loop truncations at $N=4$ and $h=0.$ See the text and the caption of Fig.~\ref{Fig:2l_local_v_full_N1} for more details on the $\phi_{\rm c}$ curves. The dashed red line corresponds to the solution of the gap equation at $\phi=0$ in the broken symmetry phase. \label{Fig:2l_local_v_full_N4}}
\end{figure}

In Fig.~\ref{Fig:2l_local_v_full_N4} a similar comparison between the full 2PI and localized results is presented in less details for the $\smash{N=4}$ case. One can see in this case that the full 2PI result is far better reproduced by the localized approximation than in the $\smash{N=1}$ case. This is possibly in connection with the appearance of approximate Goldstone modes, which, having small mass, increase the importance of the deep infrared region.\\

It is important to note that for both $N=1$ and $N=4$ an \emph{unphysical} solution to the localized equations appears in the broken phase. The reason for calling it unphysical is that it does not connect to the solution in the symmetric phase, furthermore, this solution is such that both $\bar \phi$ and $\bar M$ vanish at high temperature, as can be checked semi-analytically using the high temperature expansion (HTE).

We can trace back the existence of two broken phase solutions to the fact that the solution of the localized gap equation possesses two branches, as $\phi$ is varied.\footnote{Related to this, we mention that, sometimes, the multiplicity of solutions to the coupled system of gap and field equations is related to a first order type transition. However, one should be cautious with this type of interpretation because this depends on whether or not these various solutions originate from the same branch of the gap equation. In the Hartree approximation, the gap equation has only one branch (for values of the cutoff below the Landau scale, see \cite{Reinosa:2011ut}) and the multiplicity of solutions to the coupled gap and field equations is indeed related to a first order type transition.} In the $N=1$ case for instance, the gap equation can be rewritten as $0=g^{\mbox{\tiny $2$-loop}}_\phi(\bar M^2)$, with
\beq\label{eq:gap-2loop_for_branches}
g^{\mbox{\tiny $2$-loop}}_\phi(M^2)=g^{\mbox{\tiny $2$-loop}}_{\phi=0}(M^2)+\frac{\lambda_\star}{2}\phi^2(1-\lambda_\star {\cal B}_{\rm F}[G])\,,
\eeq  
and
\beq
g^{\mbox{\tiny $2$-loop}}_{\phi=0}(M^2)=-M^2+m^2+\frac{\lambda_\star}{2}{\cal T}_{\rm F}[G]\,.
\eeq
The function $g^{\mbox{\tiny $2$-loop}}_\phi(M^2)$ tends to $-\infty$ in the $M^2\to 0$ and $M^2\to \infty$ limits, and in between it has exactly one maximum, so for all the values of $\phi$ such that this maximum is strictly positive, there are indeed two solutions. A similar, although more involved analysis, can be done in the $\smash{N=4}$ case.  As we now discuss, the existence of multiple branches of solutions is intimately related to the loss of solutions.\\

\begin{figure}[t]
\includegraphics[width=0.44\textwidth]{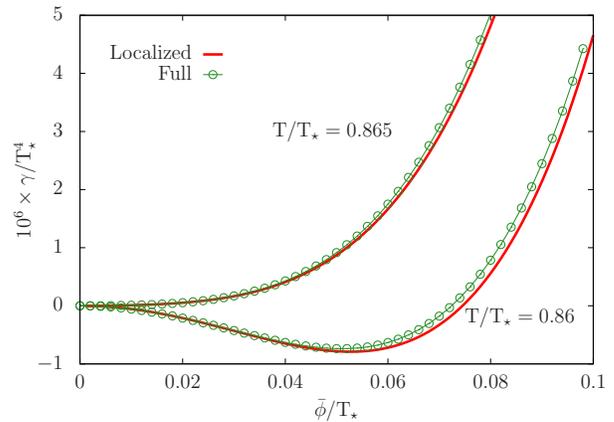}
\caption{The effective potential evaluated within the full 2PI treatment compared with the integral of the field equation with respect to the field evaluated in the localized approximation on the physical branch of the solution to the gap equation. Both temperatures are above the value $T_{\rm coal}$ appearing in Fig.~\ref{Fig:2l_local_v_full_N1} and the value of the potential at $\phi=0$ has been subtracted.\label{Fig:2l_local_v_full_N1_pot}}
\end{figure}

We note first that for large enough temperatures the two branches identified above do not cross each other and that the physical branch is defined down to $\phi=0,$ while the unphysical branch is defined only for $\phi>0$ (although it can be extended to $\phi=0$ by continuity). In the full 2PI case, to each branch denoted by $(i)$ would correspond one effective potential from the formula $\gamma_{(i)}(\phi)=\gamma[\phi,\bar G_{(i)}(\phi)]$. In the localized case, we do not have access to the functional $\gamma[\phi,G],$ but, nevertheless, we can associate a potential to each branch from the prescription
\beq
\frac{d\gamma_{(i)}}{d\phi}=\phi f^{\mbox{\tiny $2$-loop}}_\phi(\bar M^2_{(i)}(\phi))\,,
\label{eq:pot_construction}
\eeq
where $f^{\mbox{\tiny $2$-loop}}_\phi(M^2)$ is the function appearing in the right-hand side (r.h.s.) of the field equation \eqref{eq:O2lNfinField}. Using this recipe, one checks that the unphysical branch leads to a potential which has always the shape of a broken phase effective potential and the absolute minimum of this potential corresponds to the blue curve of Fig.~\ref{Fig:2l_local_v_full_N1}. The potential corresponding to the physical branch is illustrated in Fig.~\ref{Fig:2l_local_v_full_N1_pot} for two values of the temperatures above $T_{\rm coal},$ where the gap equation has solutions for any value of $\phi.$ Depending on the temperature, the potential has the shape of either a symmetric phase effective potential with a minimum at $\phi=0$, corresponding to the solid red curve in Fig.~\ref{Fig:2l_local_v_full_N1} for $T>T_{\rm c},$ or a broken phase one, with a maximum at $\phi=0$, corresponding to the dashed red curve in Fig.~\ref{Fig:2l_local_v_full_N1} for $T<T_{\rm c}$ and a nontrivial minimum corresponding to the solid red curve. One can also see in Fig.~\ref{Fig:2l_local_v_full_N1_pot} that the potential constructed by integrating the expression in the localized field equation is a good approximation of the effective potential calculated in the full 2PI case, as described in Ref.~\cite{Marko:2013lxa}.

\begin{figure}[!t]
\includegraphics[width=0.44\textwidth]{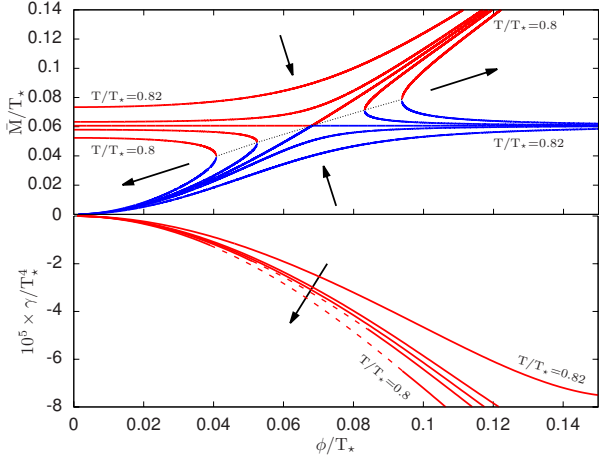}
\caption{Top panel: coalescence of the physical branch of solution to the two-loop localized gap equation (red) with the unphysical branch (blue), as the temperature is decreased. Black arrows point towards decreasing temperature values ($T/T_\star=0.82,$ 0.81, 0.807059, 0.805, and 0.8). After coalescence, a gap in $\phi$ develops where no solution to the gap equation exists (neither physical nor unphysical) and over which the potential is not defined. Bottom panel: The physical potential when varying the temperature across the temperature at which the branches coalesce (the value at $\phi=0$ is subtracted). At the considered temperatures, the potential always admits a non-trivial minimum, even though it appears at higher values of $\phi$ than the ones we show. Also the dashed curves correspond to some \emph{ad hoc} extrapolation between the two disconnected pieces of the potential below the coalescence temperature, that is $\bar T_{\rm c}<T<T_{\rm coal}$ ($T_{\rm coal}/T_\star\approx 0.807059$).\label{Fig:branch_coalescence}}
\end{figure}

\begin{figure}[!t]
\includegraphics[width=0.46\textwidth]{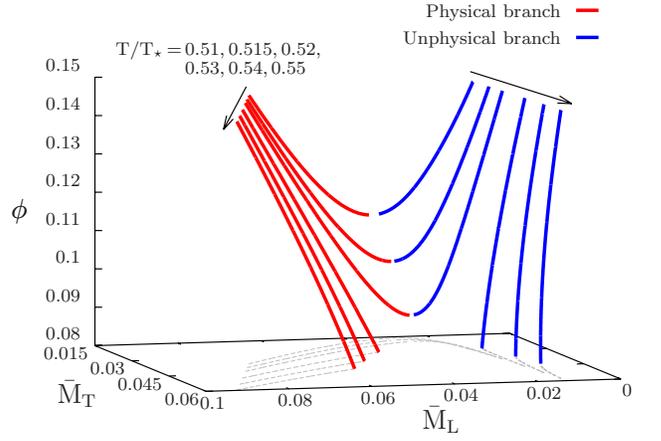}
\caption{Physical and unphysical branches of solutions to the coupled gap equations in the chiral limit of the $\smash{N=4}$ case at different temperatures (the arrows point in the direction of growing $T$). The branches can also coalesce, as in the $\smash{N=1}$ case. \label{Fig:branches_2l_N4}}
\end{figure}

If we decrease the temperature further, it can be checked that the physical and unphysical branches coalesce at a certain temperature\footnote{A discussion on the coalescence of the two branches is presented in the next subsection. There the ${\cal O}(\lambda^2)$ truncation is considered which also displays these features, for essentially the same reason as in the two-loop case.}  $T_{\rm coal},$ following the pattern shown in the top panel of Fig.~\ref{Fig:branch_coalescence}. In the $N=4$ case a similar coalescence of branches occur, as depicted in Fig.~\ref{Fig:branches_2l_N4}. Below the coalescence temperature, the branches are rearranged and a gap in $\phi$ develops where no solution to the gap equation, in particular no physical branch, exists. This means that the potential is not defined in this interval, as shown in the bottom panel of Fig.~\ref{Fig:branch_coalescence}. The gap in the range of $\phi$ visible in Fig.~\ref{Fig:branch_coalescence} can also be seen in Figs.~\ref{Fig:2l_local_v_full_N1} and \ref{Fig:2l_local_v_full_N4} where for a given $T$ it corresponds to the distance between the upper and lower branches of the curve $\phi_c$ which shows the boundary in the range of $\phi$ where the gap equation has no solution.\footnote{For comparison, the corresponding curve in the Hartree approximation is also shown (see Refs.~\cite{Reinosa:2011ut} and \cite{Marko:2013lxa}). In that case there is no bending, in line with the fact that, if the cutoff is below the scale of the Landau pole, there is no additional branch of solutions.} The appearance of such a gap in the values of $\phi$ can be problematic because, as the temperature is decreased further, it expands and could engulf some of the extrema of the potential, signaling the loss of certain (if not all) solutions to the field equation. In fact, this happens in the present case where the leftmost edge of the gap in $\phi$ reaches zero at the temperature $\bar T_{\rm c}$ corresponding to the left-end of the red dashed curve of Fig.~\ref{Fig:2l_local_v_full_N1}. Below this temperature, $\phi=0$ is not anymore a solution of the field equation. As shown in \cite{Marko:2012wc}, for parameters where both $\bar T_{\rm c}$ and $T_{\rm c}$ exist this happens for temperatures at which $\phi=0$ is already a maximum of the potential and then not anymore the physical point of the system. The latter is the absolute minimum of the potential which, in the present case, is not engulfed by the growing rightmost edge of the gap in $\phi$, so that the physical solution exists down to $T=0$, see the solid red curve of Fig.~\ref{Fig:2l_local_v_full_N1}.

To close this section, let us gather some remarks. First, although the above discussion applies to the localized equations, it is very plausible that similar multiple branches exist for the full two-loop gap equation, even though the iterative method that we use to solve the latter only allows us to access one branch. As a crosscheck, we have solved the localized equations iteratively and checked that only one branch is accessible in this way. Other methods, such as the Newton-Raphson algorithm could allow to access the unphysical branch. It is easy to convince oneself, for instance, that this is so for the localized equations in the $\smash{N=1}$ case, provided one initializes the method close enough to the solution. Second, the existence of multiple branches is probably an artifact of the truncation because in the exact theory, we expect that there should only be one branch, that is one propagator to each value of the field. Third, the presence of a second branch in the localized case is related to the fact that the function $g^{\mbox{\tiny $2$-loop}}_\phi(M^2)$ diverges negatively as $M^2\to 0$. Therefore, in the present case, the IR sensitivity is responsible for the appearance of multiple branches of solutions. Stated differently, the appearance of multiple branches signals the inability of the approximation to appropriately deal with the IR.

\subsection{Understanding the loss of solution in the ${\cal O}(\lambda^2)$ truncation from the localized approximation}\label{Sec:Compare}

Let us now consider the ${\cal O}(\lambda^2)$ case. We will see that a similar picture as the one described above applies here, but this time the physical point of the system can be engulfed by the growing gap in $\phi$. The coupled system of localized gap and field equations reads in this case
\begin{subequations}
\beq
\bml&=&m_\star^2+\frac{\lambda_\star}{2N}\left(\phi^2+{\cal T}_{\rm F}[\bgl]\right)+(N-1)\frac{\lambda_\star}{6N}{\cal T}_{\rm F}[\bgt]\nonumber\\
&&-\frac{\lambda_\star^2\phi^2}{18N^2}\left(9{\cal B}_{\rm F}[\bgl]+(N-1){\cal B}_{\rm F}[\bgt]\right)\nonumber\\
&&-\frac{\lambda_\star^2}{18N^2}\left(3{\cal S}_{\rm F}[\bgl]+(N-1){\cal S}_{\rm F}[\bgl;\bgt;\bgt]\right),\label{eq:Ol2NfinGapL}\\
\bmt&=&m_\star^2+\frac{\lambda_\star}{6N}\left(\phi^2+{\cal T}_{\rm F}[\bgl]\right)+(N+1)\frac{\lambda_\star}{6N}{\cal T}_{\rm F}[\bgt]\nonumber\\
&&-\frac{\lambda_\star^2\phi^2}{9N^2}{\cal B}_{\rm F}[\bgl;\bgt]-\frac{\lambda_\star^2}{18N^2}\left({\cal S}_{\rm F}[\bgt;\bgl;\bgl]\right.\nonumber\\
&&\left.+(N+1){\cal S}_{\rm F}[\bgt]\right),\label{eq:Ol2NfinGapT}\\
h&=&\bar\phi\left[m_\star^2+\frac{\lambda_\star}{6N}\bar\phi^2+\frac{\lambda_\star}{2N}{\cal T}_{\rm F}[\bgl]+(N-1)\frac{\lambda_\star}{6N}{\cal T}_{\rm F}[\bgt]\right.\nonumber\\
&&\left.-\frac{\lambda_\star^2}{18N^2}\left(3{\cal S}_{\rm F}[\bgl]+(N-1){\cal S}_{\rm F}[\bgl;\bgt;\bgt]\right)\right].\nonumber\\ \label{eq:Ol2NfinField}
\eeq
\end{subequations}

%
\subsubsection{$N=1$ case}

\begin{figure}[!t]
\includegraphics[width=0.48\textwidth]{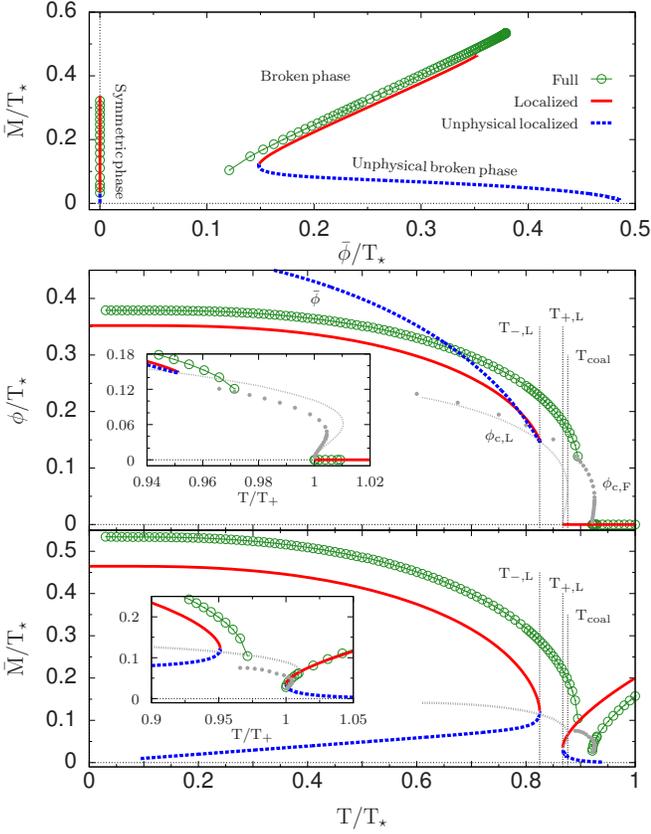}
\caption{Comparison of the localized and full 2PI results in the ${\cal O}(\lambda^2)$ truncation for $N=1$ and $h=0.$ Top panel: solution ($\bar M$) of the gap equation as function of the solution ($\bar \phi$) of the field equation. Middle panel: $\bar\phi(T),$ where it exists, together with the critical values ($\phi_{\rm c,F}$ in the full and $\phi_{\rm c,L}$ in the localized case) which limits the range of $\phi$ for which the gap equation loses its solution (see text for more details). Bottom panel: the solution of the gap equation as a function of $T.$ In the inset of both figures the temperature is scaled by $T_+,$ the temperature value where the symmetric phase gap equation loses its solution. Its value is different in the two approximations: $T_+/T_\star\approx0.921$ in the full 2PI case and $T_+/T_\star\approx0.867$ in the localized case. \label{Fig:BB_local_v_full_N1}}
\end{figure}

The solutions of the above equations in the chiral limit of the $\smash{N=1}$ case are shown in Fig.~\ref{Fig:BB_local_v_full_N1}, where they are compared with the iterative solution of the full 2PI equations. In the top panel, where the solutions are plotted parametrically as functions of the temperature, $T\to (\bar\phi(T),\bar M(T))$, we see that the presence of the setting-sun integral in the gap equation makes the local approximation less accurate compared to the two-loop case discussed in the previous subsection. In this case the solution of the gap equation is sensitive to the way the setting-sun integral is evaluated, that is with a localized or with a full 2PI propagator. 

The most important feature of the temperature evolution of the solution is that there exists a temperature region where the system of gap and field equations does not admit a solution. We use respectively $T_-$ and $T_+$ to denote the lower and upper limits of this temperature region. For $T<T_+$ the gap equation admits a solution only for field values larger than some ``critical'' value $\phi_{\rm c}$ which is estimated in the full 2PI case through a fit and computed numerically in the localized case. Above $T_+$ there is a temperature range where the $\phi_{\rm c}$ curve is bivalued, exactly as in the two-loop case (see Fig.~\ref{Fig:2l_local_v_full_N1}): when increasing the field from zero at fixed $T$ the gap equation admits a solution only up to some value of the field, above which there is no solution up to some value of the field, beyond which the gap equation has solution again. Below we investigate the loss of solution in the localized approximation which also reveals the reason for the bending of the $\phi_{\rm c}(T)$ curve shown in the middle panel of Fig.~\ref{Fig:BB_local_v_full_N1}.\\

\begin{figure}[!t]
\includegraphics[width=0.46\textwidth]{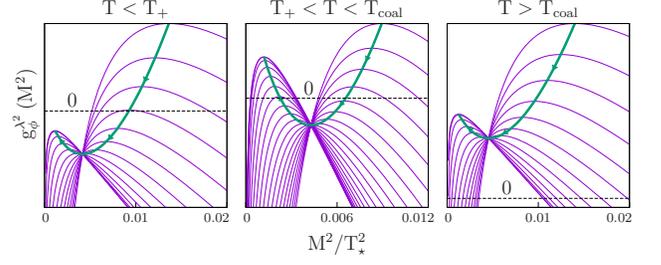}
\caption{The function $g^{\lambda^2}_\phi\left(M^2\right)$ of Eq.~\eqref{Eq:F_gap_at_phi}, which determines the ${\cal O}(\lambda^2)$ localized gap equation for $N=1$, at different values of the field and for three values of the temperature, which from left to right are: $T/T_\star=0.86,\,0.87275,$ and $0.9.$ The arrows on the curves connecting the maxima point in the direction of decreasing $\phi$ and on each plot the last $g^{\lambda^2}_\phi\left(M^2\right)$ curve along the arrows corresponds to $\phi=0$.
\label{Fig:BB_local_N1_gapeqs}}
\end{figure}

To get more insight on the disappearance of solutions, we again consider the branches of the localized gap equation which we write as $0=g^{\lambda^2}_\phi(\bar M^2)$ with
\beq
g^{\lambda^2}_\phi(M^2)=g^{\lambda^2}_{\phi=0}(M^2)+\frac{\lambda_\star}{2}\phi^2(1-\lambda_\star {\cal B}_{\rm F}[G])
\label{Eq:F_gap_at_phi}
\eeq 
and
\beq\label{eq:toto}
g^{\lambda^2}_{\phi=0}(M^2)=-M^2+m^2+\frac{\lambda_\star}{2}{\cal T}_{\rm F}[G]-\frac{\lambda_\star^2}{6}{\cal S}_{\rm F}[G]\,.
\eeq
One can plot this function and parametrize the different curves by $\phi$. This is shown in Fig.~\ref{Fig:BB_local_N1_gapeqs} for three values of the temperature, $T/T_\star=0.86,\,0.87275,$ and $0.9$. At large $T$ and $\phi$ the curves have two zeros and a maximum between them. At a fixed $T$, there exists a point $\bar M^2_\textrm{\O}$, such that all the curves with different $\phi$ go through it, that is at which the value of the gap equation is independent on $\phi.$ This specific value $\bar M^2_\textrm{\O}$ satisfies the equation $1-\lambda_\star {\cal B}_{\rm F}[\bar G_\textrm{\O}]=0,$ where $\bar G_\textrm{\O}(Q)\equiv 1/(Q^2+\bar M^2_\textrm{\O}).$ As $\phi$ decreases at fixed $T,$ the maximum of the $g^{\lambda^2}_\phi(\bar M^2)$ function shifts from large to small values of $M^2$, by passing through this special point, while the value of the function at the position of the maximum first decreases then increases, as shown by the green curve of Fig.~\ref{Fig:BB_local_N1_gapeqs}. Since all curves with different $\phi$ have to go through the point which is independent of $\phi$, the minimum of the green curve has to be the special point $\bar M^2_\textrm{\O}.$ Therefore, if the value $g^{\lambda^2}_{\phi=0}(\bar M^2_\textrm{\O})$ of the gap equation at this point is strictly positive, then two branches exist for any $\phi>0$. If $g^{\lambda^2}_{\phi=0}(\bar M^2_\textrm{\O})=0$, the two-branches coalesce at some value of $\phi$. The temperature $T_{\rm coal}$ at which this coalescence of branches occurs is determined from the equation\footnote{A similar equation defines the coalescence temperature in the two-loop case: $g^{\mbox{\tiny $2$-loop}}_{\phi=0}(\bar M^2_\textrm{\O})|_{T_{\rm coal}}=0$.} $g^{\lambda^2}_{\phi=0}(\bar M^2_\textrm{\O})|_{T_{\rm coal}}=0$. Finally, if the temperature is such that $g^{\lambda^2}_{\phi=0}(\bar M^2_\textrm{\O})<0$, there is a gap in the values of $\phi$ where the two branches cease to exist. This gap is nothing but the distance between the upper and lower branch of the grey curve in Fig.~\ref{Fig:BB_local_v_full_N1}. The gap increases as the temperature is decreased and the leftmost part of the gap reaches zero at some point. This means that below this point, $\phi=0$ is not anymore a solution of the field equation, but contrarily to the two-loop case, this happens for temperatures at which $\phi=0$ is still the absolute minimum (and unique extremum) of the potential and thus the physical point of the system. As the temperature is decreased even further, the absolute minimum reemerges by exiting the gap in $\phi$ and the physical solution reappears. This occurs for $T\le T_-.$

\begin{figure}[!t]
\includegraphics[width=0.46\textwidth]{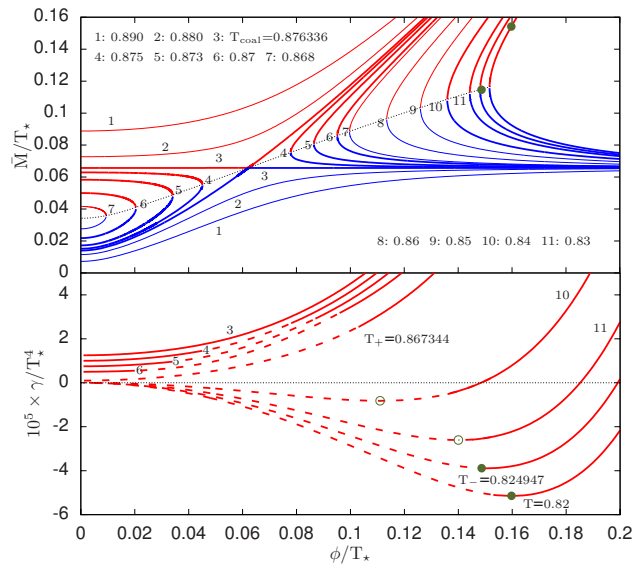}
\caption{Top panel: coalescence of the physical branch of solution to the ${\cal O}(\lambda^2)$ localized gap equation (red) with the unphysical branch (blue) as the temperature is decreased. The temperature values are in units of $T_\star.$ After coalescence, a gap in $\phi$ develops where no solution to the gap equation exists (neither physical nor unphysical) and over which the potential is not defined. Bottom panel: The physical potential for the temperature values corresponding to the thick lines in the top panel and for $T_+$. The value of the potential at $\phi=0$ is subtracted, but for the symmetric phase potentials an arbitrary constant is added for the sake of the presentation. The dashed curves correspond to some \emph{ad hoc} extrapolation as detailed in the text. The blob indicates the nontrivial minimum of the potential (solution of the field equation).\label{Fig:branch_coalescence_Ol2}}
\end{figure}

The behavior of the two branches of solutions to the gap equation is illustrated in Fig.~\ref{Fig:branch_coalescence_Ol2} for temperatures around $T_{\rm coal}.$ There, we also present the potential constructed by integrating the expression in the field equation, as discussed in the two-loop case, for temperatures around $T_{\rm coal}$ and $T_-$. When constructing the potential in the range $T_+<T<T_{\rm coal},$ for those values of the field where the gap equation has no solution, we used  the value of $\bar M$ which can be read from the dashed curve appearing in the top panel of Fig.~\ref{Fig:branch_coalescence_Ol2}. This curve connects the endpoints of branches of solutions which occurs at various values of $T$\footnote{We could choose also a straight line connecting the endpoints of the branches appearing at the value of the temperature for which we draw the potential.} and the corresponding part of the potential is drawn with a dashed line. This procedure fails to give a potential with expected shape for temperatures below $T_-,$ where there is no solution to the gap equation around $\phi=0.$ In this case the dashed part of the potential represents just the value for $\phi<\phi_{\rm c}$ of a cubic polynomial in $\phi^2$, fitted to the potential constructed for $\phi>\phi_{\rm c}$.\\

To close this section, let us mention that it is simple to understand why a critical temperature cannot be reached in the localized approximation. Let us proceed by contradiction. If there were a $T_{\rm c}$, at which $\bar M=0$ and $\bar\phi=0$, we would have
\beq
0=m^2+\frac{\lambda_\star}{2}{\cal T}_{\rm F}^{T_{\rm c}}[\bar G_{\rm c}]-\frac{\lambda_\star}{6}{\cal S}_{\rm F}^{T_{\rm c}}[\bar G_{\rm c}],
\eeq
but this equation does not make sense because the last term is infinite while the others are finite, as according to \cite{Parwani:1991gq}, the finite setting-sun integral diverges for $\bar M\to 0$ as ${\cal S}_{\rm F}[\bar G]\sim -T^2\log\frac{\bar M^2}{T^2}$. Moreover, the use of the HTE helps understanding why there is a gap in temperature where the solution is lost if the mass decreases enough (so that the HTE can be used) when the temperature is decreased (resp. increased) from the symmetric (resp. broken phase); for a similar argument in the full case see Sec.~\ref{sec:anomalous}. First let us decrease the temperature from the symmetric phase. In this case we only have to discuss the gap equation at $\phi=0,$  that is $0=g^{\lambda^2}_{\phi=0}(\bar M^2)$, with $g^{\lambda^2}_{\phi=0}(M^2)$ given in (\ref{eq:toto}). As $\bar M/T$ becomes smaller and smaller, the  integral ${\cal T}_{\rm F}[\bar G]$ goes to a $T$ dependent constant, while ${\cal S}_{\rm F}[\bar G]$ diverges logarithmically. Therefore, we see that the gap equation at $\phi=0$ cannot have a solution below some temperature. Approaching from the broken phase one uses the gap equation $0=g^{\lambda^2}_\phi(\bar M^2)$ in the field equation \eqref{eq:Ol2NfinField} (at $N=1$ and $h=0$) to arrive at
\beq
\bar\phi^2 = -\frac{6\bar M^2}{3\lambda_\star^2{\cal B}_{\rm F}[\bar G]-2\lambda_\star},
\label{eq:brokenphi_N1_BB}
\eeq
which, as $\bar M$ becomes small enough, eventually turns negative when ${\cal B}_{\rm F}[\bar G]\sim T/\bar M$ overgrows $2/(3\lambda_\star).$ This shows that the solution will be lost above some temperature $T_-$, in line with our observations.\\

\begin{figure}[!t]
\includegraphics[width=0.46\textwidth]{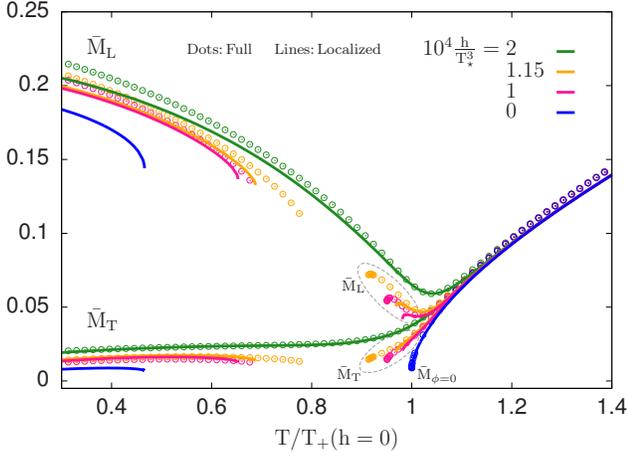}
\caption{The temperature dependence of the $\bar M_{\rm L}$ and $\bar M_{\rm T}$ gap masses at the solution of the field equation for several values of the external source $h$. Both in the full 2PI (dots) and localized (lines) ${\cal O}(\lambda^2)$ truncations, an iterative solution cannot be found for small enough $h$ in some temperature range. See the text for explanation.
\label{Fig:BB_local_v_full_N4}}
\end{figure}

\subsubsection{$N=4$ case}

The picture arising in the $\smash{N=4}$ case is very similar to the one obtained in the one-component case. While both a symmetric phase and a broken phase exist, the solution of the full ${\cal O}(\lambda^2)$ truncation is lost in an intermediate region of temperature around the would-be critical temperature. By introducing a non-zero external source $h$ (which is needed to generate a physical pion mass) one can study how the temperature region without solution first shrinks, then disappears completely with increasing values of $h.$ A comparison of the full and localized ${\cal O}(\lambda^2)$ results at $N=4$ is shown in Fig.~\ref{Fig:BB_local_v_full_N4} for several values of $h$. We could not find an iterative solution to the full 2PI equations in the chiral limit, where convergence of the under-relaxation method would most probably require $\alpha\to0$ in \eqref{Eq:under-relax} as a result of the infrared divergence which shrinks the domain of convergence to zero.

\begin{figure}[!t]
\includegraphics[width=0.46\textwidth]{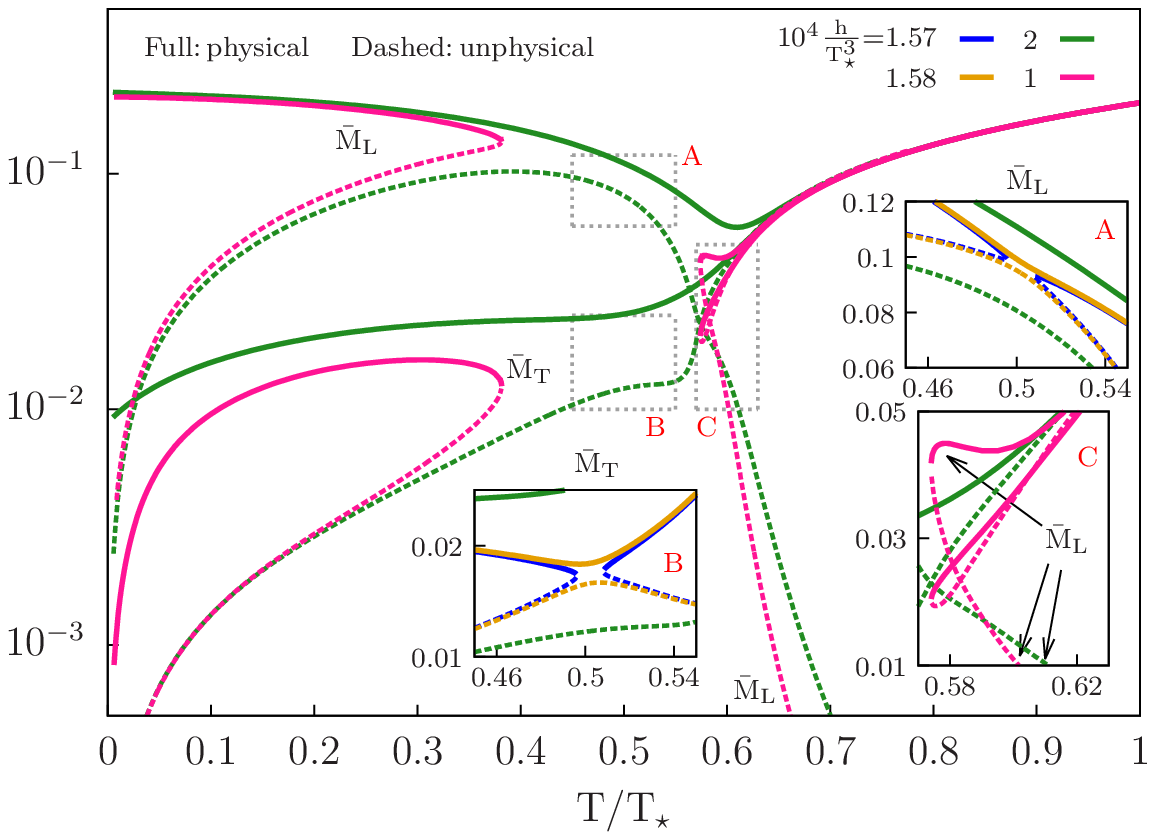}
\caption{Solution of the localized equations of the ${\cal O}(\lambda^2)$ truncation at $N=4.$  Only gap masses are shown around the critical value of $h$ where the gap in the temperature evolution of the solution appear. Notice the logarithmic scale of the y-axis. The insets have linear y-axis and zoom into some regions of interest. Compared to the main plot, curves with values of $h$ closer to the critical value are added in the insets A and B. \label{Fig:BB_local_allSols_N4}}
\end{figure}

The disappearance of the temperature gap in the solution above a certain value of $h$ is illustrated in the localized  ${\cal O}(\lambda^2)$ truncation in Fig.~\ref{Fig:BB_local_allSols_N4}. We also show there the unphysical solutions, which are present in the localized approximation. Notice, that these solutions do not go away at large values of $h$, they merely disconnect from the physical ones, allowing for a continuous temperature evolution of both physical and unphysical solutions. These plots illustrate once again that the disappearance of physical solutions originates in the coalescence of physical branches of solutions with unphysical ones.

\begin{figure}[!b]
\includegraphics[width=0.46\textwidth]{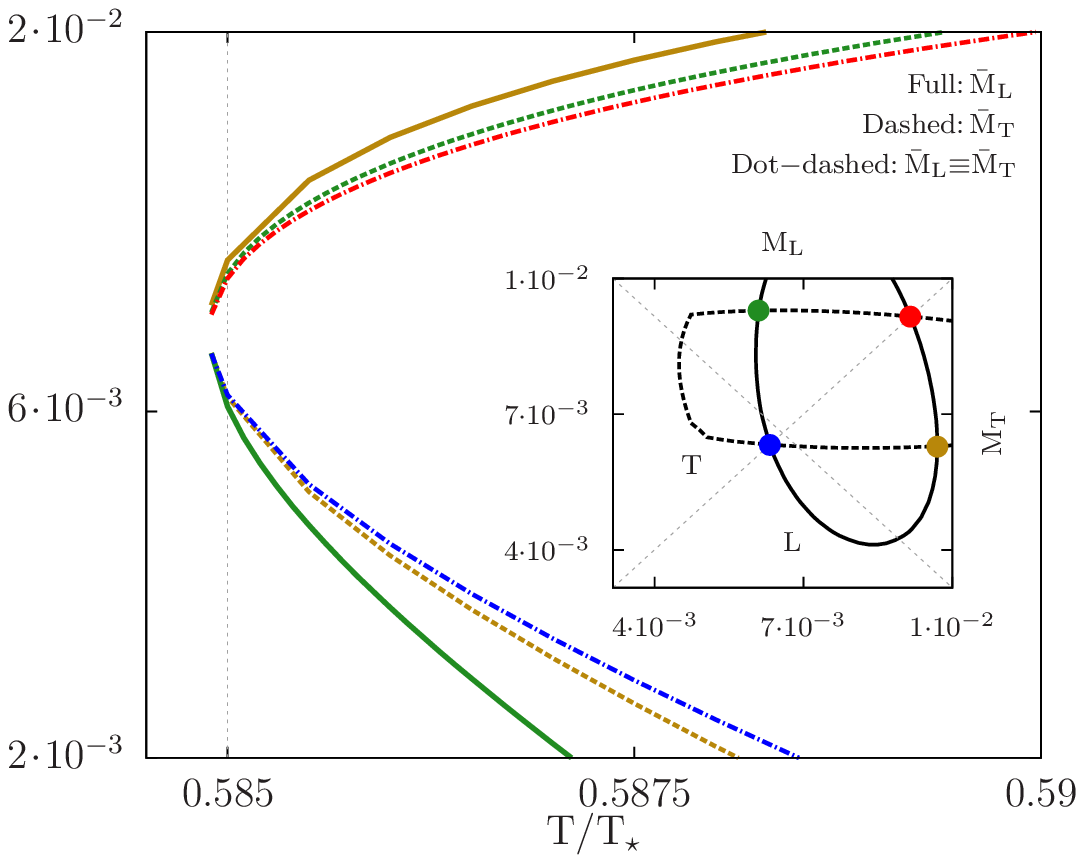}
\caption{Solutions of the gap equation for $\bar\phi=0$ in the localized approximation of the ${\cal O}(\lambda^2)$ truncation at $N=4.$ The only physical solution is the red one (upper dot-dashed), the other three colors represent unphysical solutions. The inset shows at $T/T_\star=0.585$ the curves in the $M_{\rm L}{\rm-}M_{\rm T}$ plane along which the longitudinal (full) or the transverse (dashed) gap equation is satisfied. The intersection points of these curves specify all the possible solutions of the coupled gap equations, so that each colored blob corresponds to one point on the same colored curve of the main plot. 
\label{Fig:Crazy}}
\end{figure}

We conclude this section by mentioning that, in the chiral limit ($h=0$), the localized approximation exhibits some interesting properties at high temperature, where there exist three unphysical solutions, in addition to the physical one. From these unphysical solutions only one is such that $\bar M_{\rm T}=\bar M_{\rm L}$, while for the other two the $O(N)$ symmetry is broken, although $\phi=0$.\footnote{This feature could be interesting if it would happen for the physical solution because it would signal a more subtle breaking of the $O(N)$-symmetry.} As the temperature is lowered, all four solutions cease to exist at the same temperature, $T_+$. This is presented in Fig.~\ref{Fig:Crazy}, where all the four solutions are shown, together with zero level contour lines in the $M_{\rm L}{\rm-}M_{\rm T}$ plane showing at $T/T_\star=0.585$ the mass values where the transverse and longitudinal gap equations are satisfied for $\phi = 0.$ Both curves are closed and their points of intersection specify all the possible solutions. A common property of the unphysical solutions is that at least one of the masses goes to zero as the temperature grows.

\section{Final remarks}\label{sec:remarks}

\subsection{Anomalous dimension \label{sec:anomalous}}
It is pretty clear in the full non-local ${\cal O}(\lambda^2)$ truncation that the existence of a critical temperature in the chiral limit would require the propagator to become anomalous. To see this we shall concentrate on the $K=0$ mode and following our notational convention we shall omit writing the momentum dependence for quantities with vanishing external momentum.

Suppose indeed that a critical temperature $T_{\rm c}$ (such that $\hat M^2_{\phi=0,T_{\rm c}}=\bar M^2_{\phi=0,T_{\rm c}}$) were to exist. At $T=T_{\rm c}$, we would have
\beq
0=m^2_0+\frac{\lambda^{(NA+2B)}_0}{6N}{\cal T}_{T_{\rm c}}[\bar G_{\rm c}]-\frac{(N+2)\lambda_\star^2}{18N^2}{\cal S}_{T_{\rm c}}[\bar G_{\rm c}]\,,\quad
\eeq
with $G_{\rm c}^{-1}(K=0)=0$. But the previous equation is absurd if the propagator is not anomalous, that is if $G_{\rm c}(0,k)\sim 1/k^2$ as $k\to 0$. The fact that we do not obtain numerically a $T_{\rm c}$ value in the ${\cal O}(\lambda^2)$ truncation could then indicate that the equation is not able to generate such an anomalous propagator and that the loss of solution has an infrared origin. The very presence of a whole interval where the solution disappears might be understood from the incapacity of the gap equation to generate an anomalous behavior. In fact, the gap equation at $K=0$ in the symmetric phase reads
\beq
\bar M^2=m^2_0+\frac{\lambda^{(NA+2B)}_0}{6N}{\cal T}[\bar G]-\frac{(N+2)\lambda_\star^2}{18N^2}{\cal S}[\bar G]\,.
\eeq
If we assume that the mass becomes smaller and smaller when decreasing $T$, and if no anomalous dimension is generated, there is a temperature $T_+$ below which this equation is absurd since the left-hand side (l.h.s.) is positive while the r.h.s. is negative. A similar line of thought can be done in the broken phase. Using the longitudinal gap equation \eqref{eq:bml} and exploiting the expression \eqref{Eq:l4_explicit} for the bare coupling $\lambda_4$ in the field equation \eqref{eq:bphi}, one finds that the broken phase configuration obeys the equation
\beq
\bar M^2_{\rm L}=\frac{\lambda_\star}{3N}\bar\phi^2\left[1-\frac{\lambda_\star}{6N}\Big(9{\cal B}_{\rm F}[\bgl]+(N-1){\cal B}_{\rm F}[\bgt]\Big)\right].\nonumber\\
\eeq
If we assume that the longitudinal mass becomes smaller and smaller when increasing $T$, and if no anomalous dimension is generated, there is a temperature $T_-$ above which this equation is absurd since the l.h.s. is positive while the r.h.s. is negative.

 \subsection{Limits of localization}\label{sec:lim_loc}
We have emphasized and also exemplified that the localization of a self-consistent propagator equation can lead to a good approximation of the solution of the original equation only if the gap mass is not vanishing in the deep IR. It is relatively easy to write a self-consistent equation for which the localization fails, as the solution of the original equation in the deep IR is massless and characterized by an anomalous dimension $\eta.$ In order to show this, we consider the equation
\beq
G^{-1}(k)=k^2+m^2-c\int_k^\mu d p\, p\,G(p)\,,
\label{Eq:example}
\eeq
with $m^2>0$ and $c>0.$

It is obvious that the localized version of Eq.~\eqref{Eq:example} written for a propagator of the form $G(k)=1/(k^2+M^2)$ by setting $k=0$ in the lower limit of the integral cannot lead to an anomalous behavior of the solution in the IR. Moreover, it can be shown that the solution of the localized equation is lost as $m^2$ is decreased. After performing the integral the explicit localized equation is 
\beq
0=-M^2+m^2-\frac{c}{2}\ln\frac{\mu^2+M^2}{M^2}.
\label{Eq:example_loc}
\eeq 
Since the first derivative of the r.h.s. of \eqref{Eq:example_loc} with respect to $M^2$ decreases from $\infty$ to $-1$ in the $M^2\in[0,\infty)$ range, the r.h.s. of \eqref{Eq:example_loc} is a concave function of $M^2$ having one maximum. The value of the function at the maximum decreases with $m^2,$ leading to the loss of solution at the value of $m^2$ for which the value of the function at the maximum vanishes.

The solution of the original equation \eqref{Eq:example} with a momentum dependent integral can be obtained by using separation of variables in the ordinary differential equation obtained by differentiating \eqref{Eq:example_loc} with respect to $k.$ The solution can be written as
\beq
k^2 = G^{-1}(k) - \frac{c}{2}\ln\left(2\frac{G^{-1}(k)}{c}+1\right) + g(m^2),
\label{Eq:example_sol}
\eeq  
where $g(m^2)=-m^2+(c/2)\ln\left(1+2(\mu^2+m^2)/c\right).$ This function $g(m^2)$ is such that $g(0)>0$, $g(\infty)=-\infty$ and $g'(m^2)<0$, therefore there exits $m^2_{\rm c}>0$ such that $g(m_{\rm c}^2)=0.$ Then, we see from \eqref{Eq:example_sol} that the solution at $m^2_{\rm c}$ is compatible with $G^{-1}(0)=0.$ To see the dependence of the propagator in the IR we expand the logarithm and obtain $G^{-1}(k)\sim c^{1/2} k,$ that is an anomalous propagator $G(k)\sim k^{\eta-2},$ with anomalous dimension $\eta=1.$

The equation \eqref{Eq:example} can be solved iteratively using a discretization of the momenta and performing the integral numerically. The solution \eqref{Eq:example_sol} is reproduced and as $m^2$ approaches $m_{\rm c}^2$ there is a region of the momentum in the IR where, with the exception of the first few values of the momentum, the propagator shows the anomalous behavior.

\subsection{On the solution of the Symmetry Improved (SI) 2PI equations}\label{sec:SI2PI}

We mentioned in the Introduction that a possible way to reconcile the $\Phi$-derivable expansion scheme of the $O(N)$ model with the symmetry constraint $\bar\phi\bar M^2_{\rm T}=h$ is to follow the interesting approach recently put forward by Pilaftsis and Teresi in \cite{Pilaftsis:2013xna}. In the chiral limit ($h\to 0$) this procedure boils down to replacing the field equation by
\[
\bar\phi \bar M^2_{\rm T}(K=0) = 0,
\]
while leaving unchanged the two gap equations. In the broken symmetry phase an expression for $\bar\phi$ can then be obtained from $\bar M^2_{\rm T}(K=0) = 0$.

In light of the results of the present paper, a word of caution is in order, however. Two related questions arise here. First, we have shown that we could have a forbidden region of $\phi$ where the gap equations do not have a solution, and then the question is whether $\bar\phi$ obtained above belongs to this region or not. Second, numerically we do not have access to $k=0,$ as with our discretization scheme the smallest momentum is $\kappa=\Lambda/N_s,$ which introduces a natural infrared regulator. In particular, the constraint is imposed in practice as $\bar\phi\bar M^2_{\rm T}(\omega_n=0,k=\kappa)=0$. If $N_s$ is not too large, the infrared sensitivity of the equations to the transverse gap mass is tamed and the equations display solutions. However, as $N_s$ is increased, the IR sensitivity is enhanced and, depending on the truncation, some loss of solution may be observed. The question is then what happens with the solution of the equations, as we increase $N_s.$

We think that the question of a forbidden region in the $T-\phi$ plane in which the gap equations do not admit a solution has to be addressed in the SI2PI approach the same way as in the usual 2PI formalism. In a general truncation of the 2PI effective action, the existence of such a region and whether a solution with a vanishing transverse gap mass can be found outside this region is an open question. However, we have shown within our local approximation that in the two-loop and ${\cal O}(\lambda^2)$ truncations this region exists and that outside this region $\bar M_{\rm T}\neq0$.\footnote{Actually, in the localized two-loop approximation, apart from the $(\bar T_{\rm c},0)$ point of the $T-\phi$ plane, where $\phi_{\rm c}\to 0$ and $\bar M_{\rm T}\to 0$, $\bar M_{\rm T}$ can be small only in the vicinity of $(0,\phi_{\rm c}),$ without decreasing to zero.} We also see strong numerical evidence that the full 2PI solution shares this property with its localized version. This seems to indicate that the symmetry improvement fails in this case.

\begin{figure}[!t]
\includegraphics[width=0.46\textwidth]{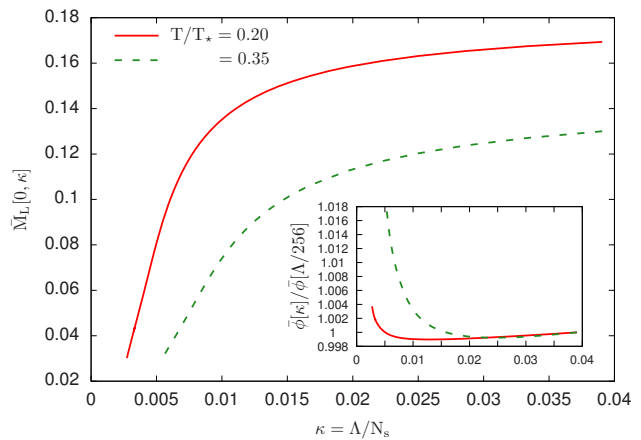}
\caption{Solution of the symmetry improved 2PI equations obtained in the two-loop truncation of the $O(N=4)$-model in the chiral limit. We show the longitudinal gap mass at the smallest available momentum $\kappa,$ as a function of $\kappa.$ The corresponding value of the condensate is shown in the inset, scaled by the value obtained at the smallest $N_s$ used. A loss of solution is observed at small values of $\kappa.$ The cutoff is $\Lambda/T_\star=10$ and the parameters are $m^2_\star/T^2_\star=0.04$ and $\lambda_\star=3.$ \label{Fig:SI2PI}}
\end{figure}

The same conclusion seems to emerge from investigating the behavior of the solution to the symmetry improved two-loop truncations as $N_s$ is increased. We concentrate on the broken symmetry phase and make use of the constraint $\bar M^2_{\rm T}(0,\kappa) = 0$ in the transverse gap equation to obtain
\[
\bar M^2_{\rm T}(K) = -\frac{\lambda_\star^2\phi^2}{9 N^2}
\left[
{\cal B}[\bar G_{\rm L},\bar G_{\rm T}](K) - 
{\cal B}[\bar G_{\rm L},\bar G_{\rm T}]\left(0,\kappa\right)
\right].
\]
By construction, $\bar M_{\rm T}(K)$ vanishes for $\omega_n=0$ and $k=\Lambda/N_s,$ which means that the integral ${\cal B}[\bar G_{\rm T}](K)$ appearing in the equation for $\bar M_{\rm L}[K]$ (see \eqref{eq:bml} and disregard the setting-sun integrals) is IR sensitive and grows as $K$ is lowered. This in turn will make the solution $\bar M_{\rm L}(0,\kappa)$ smaller and smaller as $N_s$ is increased (due to the minus sign in front of the integral), which will make ${\cal B}[\bar G_{\rm L}](0,\kappa)$ grow as well with $N_s$. Since in case of a finite $\bar\phi$ nothing can stop this positive feedback, the expectation is that we run into problems with increasing $N_s.$ This is illustrated in Fig.~\ref{Fig:SI2PI}, where one shows that the iterative solution of the model in the chiral limit of the $N=4$ case is lost in the SI2PI approach as one decreases $\kappa=\Lambda/N_s$ at fixed cutoff $\Lambda/T_\star=10.$ At a higher temperature the solution is lost for a larger value of $\kappa$ (smaller value of $N_s$).

\section{Conclusions}\label{sec:conclusions}

We have investigated the solution of the $O(N)$-model in the ${\cal O}(\lambda^2)$ truncation of the 2PI effective action. A zero temperature parametrization of the model for $N=4$ based on curvature masses revealed that for parameters where relatively high values of the longitudinal curvature mass of order $400$~MeV is attained, the scale of the Landau scale is lower compared to what was obtained in a truncation at two-loop level. 

As an interesting feature of the model in the  ${\cal O}(\lambda^2)$ truncation we found that by lowering the value of the external field $h$ the solution is lost in some temperature range. We have thoroughly investigated this feature in the chiral limit by introducing a localization of the self-consistent propagator and field equations. Our analysis revealed that this loss of solution is due to the appearance of a second (unphysical) solution to the gap equation which can coalesce with the physical branch leading to the opening of a gap in the possible values of the field, where the gap equation has no solution. The appearance of this second branch is a result of the IR sensitivity of the gap equation. Contrary to the two-loop case, where a similar situation occurs but the solution $\bar\phi$ of the field equation never enters the problematic field region, in the  ${\cal O}(\lambda^2)$ truncation this region can engulf the value of $\bar\phi$, as the temperature or the external field is varied

The problem that we have identified here in the case of the ${\cal O}(\lambda^2)$ $\Phi$-derivable approximation is more general and applies potentially to any set of self-consistent equations, such that, in some limit, some modes become light. This poses an important challenge to 2PI approximations because it is \emph{a priori} unclear whether a given approximation has the capability to tame the potential IR sensitivities, and the corresponding loss of solutions, that the appearance of light modes can trigger. The taming of IR sensitivities could occur in two different, but maybe not completely unrelated, ways: either the approximation has the capability to generate an anomalous dimension but this does not seem to be the case in the approximation considered here (at least not in the renormalization scheme that we considered), or it might be necessary to resum potential infrared divergent diagrams in the $\Phi$-derivable functional. Both these features appear in the 2PI $1/N$ expansion at next-to-leading order and therefore the model deserves to be explored in this approximation in order to further investigate the fate of the IR sensitivity arising close to the phase transition.

For the sake of completeness, we mention that our results concerning the loss of solutions have been obtained in a particular renormalization scheme where the system is assumed to be in the symmetric phase at the temperature $T_\star$ where the renormalization conditions are imposed. There exist other schemes which allow one to renormalize the system directly in the broken phase and it would be interesting to see whether our findings extend to these cases as well. Again a necessary condition for this scheme to allow us to access a possible critical region is that an anomalous dimension is generated. A scheme where this could happen is a particular limit of the one we have used, in which one assumes that there is a temperature at which the zero-momentum mass is equal to $0$ and to choose $T_\star$ equal to this temperature (and thus $m_\star=0$). Such a scheme however cannot allow us to decide on the second nature of the transition. It could at best allow us to illustrate that the hypothesis of a second order phase transition is consistent.

\acknowledgments{This work was supported by the Hungarian-French collaboration program T{\'E}T\_11-2-2012 (PHC Balaton No. 27850RB). Zs.\ Sz. would like to thank Gergely Fej\H{o}s for sharing during a common work his notes on the renormalization of the $O(N)$ model presented in Ref.~\cite{Berges:2005hc}, which proved helpful when writing App.~\ref{app:Ol2ren}. G.\ M. and Zs.\ Sz. were supported by the Hungarian Scientific Research Fund (OTKA) under Contract No. K104292.}\\

\appendix

\section{Renormalization of the ${\cal O}(\lambda^2)$ truncation\label{app:Ol2ren}}
In this Appendix we summarize some properties of the 4-point functions and present the procedure we use to fix the counterterms. 

\subsection{Four-point functions \label{app:4pFn}}
It was shown in \cite{Berges:2005hc} that a possible definition of the four-point function is through the Bethe-Salpeter equation
\beq
\bar V_{ab,cd}(K,P)&=&\bar \Lambda_{ab,cd}(K,P)-\frac{1}{2}\int_Q^T \bar V_{ab,a'b'}(K,Q) \times \nonumber\\
&&\quad\bar G_{a'c'}(Q)\bar G_{b'd'}(Q) \bar\Lambda_{c'd',cd}(Q,P),\ \ 
\label{Eq:BS_bV}
\eeq
with the kernel
\beq
\bar\Lambda_{ab,cd}(K,P)=4\frac{\delta^2\gamma_\textnormal{int}[\phi,G]}{\delta G_{ab}(K) \delta G_{cd}(P)}\bigg|_{\bar G}\,,
\eeq
which admits the following decomposition in terms of invariant tensors:
\beq
\bar\Lambda_{ab,cd}(K,P)&=&\delta_{ab}\delta_{cd}\bar\Lambda^A(K,P)+\delta_{ac}\delta_{bd}\bar\Lambda^B(K,P)\nonumber\\
&+&\delta_{ad}\delta_{bc}\bar\Lambda^B(K,-P).
\label{Eq:bL_dec_F}
\eeq
The components obey the following properties:
\begin{widetext}
\beq
\bar\Lambda^A(K,P)=\bar\Lambda^A(K,-P)=\bar\Lambda^A(-K,-P),\qquad 
\bar\Lambda^B(K,P)=\bar\Lambda^B(-K,-P), \quad \bar\Lambda^B(-K,P)=\bar\Lambda^B(K,-P).
\label{Eq:bL-AB_prop}
\eeq
Plugging the decomposition \eqref{Eq:bL_dec_F} and a similar one for $\bar V_{ab,cd}(K,P)$ into \eqref{Eq:BS_bV}, one arrives using \eqref{Eq:bL-AB_prop} at a set of coupled Bethe-Salpether equations for $\bar V^A(K,P)$ and $\bar V^B(K,P).$ In order to decouple them it is convenient to introduce the combinations
\beq
\bar \Lambda^S(K,P)=N \bar \Lambda^A(K,P) + \bar \Lambda^B(K,P) + \bar \Lambda^B(K,-P),\quad
\bar V^S(K,P)=N \bar V^A(K,P) + \bar V^B(K,P) + \bar V^B(K,-P),\ \ 
\eeq
which then gives
\beq
\bar V^B(K,P)&=&\bar\Lambda^B(K,P)-\int_Q^T\bar\Lambda^B(K,Q)\bar G^2(Q) \bar V^B(Q,P),\label{Eq:bVB}\\
\bar V^S(K,P)&=&\bar\Lambda^S(K,P)-\frac{1}{2}\int_Q^T\bar\Lambda^S(K,Q)\bar G^2(Q) \bar V^S(Q,P).\label{Eq:bVS}
\eeq
The above equations are needed for renormalization at $\phi=0$ and in the concrete case of the skeleton order ${\cal O}(\lambda^2)$ truncation the kernels are:
\begin{subequations}
\beq
\bar\Lambda^A_{\phi=0}(K,P)&=&\frac{\lambda_0^{(A)}}{3N}-\frac{\lambda_\star^2}{9N^2}\big[{\cal B}[\bar G_{\phi=0}](K+P)+{\cal B}[\bar G_{\phi=0}](K-P)\big], \label{Eq:bLA}\\
\bar\Lambda^B_{\phi=0}(K,P)&=&\frac{\lambda_0^{(B)}}{3N}-\frac{\lambda_\star^2}{18N^2}\big[(N+4){\cal B}[\bar G_{\phi=0}](K+P)+2{\cal B}[\bar G_{\phi=0}](K-P)\big], \label{Eq:bLB}\\
\bar\Lambda^S_{\phi=0}(K,P)&=&\frac{\lambda_0^{(NA+2B)}}{3N}-\frac{\lambda_\star^2}{6N^2}(N+2)\big[{\cal B}[\bar G_{\phi=0}](K+P)+{\cal B}[\bar G_{\phi=0}](K-P)\big]. \label{Eq:bLS}
\eeq
\end{subequations}
\end{widetext}
As was shown in \cite{Berges:2005hc}, due to the relation \eqref{Eq:spec_rel_trunc} the four-point function
\beq
\hat V^{\phi=0}_{abcd}=\frac{\delta^4\gamma(\phi)}{\delta \phi_a \delta\phi_b \delta\phi_c \delta\phi_d}\Big|_{\phi=0}
\eeq
satisfies
\beq
\hat V^{\phi=0}_{abcd} = \hat\Lambda^{\phi=0}_{abcd}&+&\bar V^{\phi=0}_{ab,cd}(0,0)-\bar \Lambda_{ab,cd}(0,0) \nonumber\\
&+& \bar V^{\phi=0}_{bc,ad}(0,0)-\bar \Lambda_{bc,ad}(0,0) \nonumber\\ 
&+& \bar V^{\phi=0}_{bd,ac}(0,0)-\bar \Lambda_{bd,ac}(0,0),
\label{Eq:hatV_spec_trunc}
\eeq
where in the skeleton order ${\cal O}(\lambda^2)$ truncation $\hat\Lambda^{\phi=0}_{abcd}=\frac{\delta^4\gamma_\textrm{int}[\phi,G]}{\delta \phi_a \delta\phi_b \delta\phi_c \delta\phi_d}\Big|_{\phi=0}=\frac{\lambda_4}{3N}\big(\delta_{ab}\delta_{cd}+\delta_{ac}\delta_{bd}+\delta_{ad}\delta_{bc}\big).$ Then,  using the decomposition $\hat V^{\phi=0}_{abcd}=\hat V_{\phi=0}\big(\delta_{ab}\delta_{cd}+\delta_{ac}\delta_{bd}+\delta_{ad}\delta_{bc}\big)$ and the relation \eqref{Eq:bL_dec_F} for $\phi=0$ with a similar one for $\bar V_{ab,cd}^{\phi=0}(0,0),$  one obtains from \eqref{Eq:hatV_spec_trunc} 
\beq
\hat V_{\phi=0}=\frac{\lambda_4}{3N}&+&\bar V^A_{\phi=0}(0,0) - \bar\Lambda^A_{\phi=0}(0,0) \nonumber\\
&+& 2\big(\bar V^B_{\phi=0}(0,0) - \bar\Lambda^B_{\phi=0}(0,0)\big).
\label{Eq:hatV_spec_trunc2}
\eeq

\subsection{Determination of $\delta Z_\tau,$ $\delta Z_s,$ and $\delta m_0^2$}
The propagator equation at $T_\star$ and vanishing field is
\beq
G_\star^{-1}(K_\star)&=&Z_\tau\omega_\star^2+Z_s\k^2+m_0^2+\frac{\lambda_0^{(NA+2B)}}{6N}{\cal T}_\star[G_\star] \nonumber\\
&-&\frac{N+2}{18N^2}\lambda_\star^2\int_{Q_\star}^{T_\star}{\cal B}_\star(Q_\star) G_\star(K_\star-Q_\star)\,, 
\label{Eq:Gstar}
\eeq
where we introduced the shorthand ${\cal B}_\star(Q_\star)\equiv{\cal B}_\star[G_\star](Q_\star).$ In order to determine the counterterms $\delta Z_\tau,$ $\delta Z_s,$ and $\delta m_0^2$ we impose the following renormalization (and consistency) conditions \footnote{\eqref{Eq:ren_cond_Zt} and \eqref{Eq:ren_cond_m2} are renormalization conditions, while \eqref{Eq:ren_cond_Zs} is considered a consistency condition.}:
\begin{subequations}
\beq
\label{Eq:ren_cond_Zt}
&&G_\star^{-1}(\hat K_\star)-G_\star^{-1}(\tilde K_\star)=\hat\omega_\star^2-\tilde\omega_\star^2,\\
\label{Eq:ren_cond_Zs}
&&G_\star^{-1}(\bar K_\star)-G_\star^{-1}(\tilde K_\star)=\hat\k^2-\tilde\k^2,\\
\label{Eq:ren_cond_m2}
&&G_\star^{-1}(\tilde K_\star)=\tilde K_\star^2+m_\star^2,
\eeq
\end{subequations}
where $\tilde K_\star=(i\tilde\omega_\star, \tilde \k),$ $\hat K_\star=(i\hat\omega_\star,\tilde \k),$ and $\bar K_\star=(i\tilde\omega_\star,\hat \k)$ with $\tilde\omega_\star=\Delta\omega=2\pi T_\star,$  $\hat\omega_\star=2 \Delta\omega,$ $|\tilde\k|=\Delta k=\Lambda/N_s,$ and $|\hat\k|=2 \Delta k.$ Here $\Delta\omega$ and $\Delta k$ are the lattice spacing in frequency and momentum space after exploiting the rotational invariance in momentum space. The above conditions give:
\begin{subequations}
\label{Eq:dZt-dZs-dm02}
\beq
\delta Z_\tau &=& \frac{ \lambda_\star^2(N+2)/N^2}{18(\tilde\omega^2_\star-\hat\omega^2_\star)}\int_{Q_\star}^{T_\star}{\cal B}_\star(Q_\star) \delta G_\star(Q_\star;\tilde K_\star,\hat K_\star),\qquad\quad\\
\delta Z_s &=& \frac{ \lambda_\star^2(N+2)/N^2}{18(\tilde\k^2-\hat\k^2)}\int_{Q_\star}^{T_\star}{\cal B}_\star(Q_\star) \delta G_\star(Q_\star;\tilde K_\star,\bar K_\star),\qquad\quad\\
\delta m_0^2 &=& -\delta Z_\tau\tilde\omega_\star^2-\delta Z_s\tilde\k^2 - \frac{\lambda_0^{(NA+2B)}}{6N}{\cal T}_\star[G_\star] \nonumber\\
&& + \frac{(N+2)\lambda_\star^2}{18N^2}\int_{Q_\star}^{T_\star}{\cal B}_\star(Q_\star)G_\star(\tilde K_\star-Q_\star),
\eeq
\end{subequations}
with the shorthand $\delta G_\star(Q_\star;K_\star,P_\star)=G_\star(K_\star-Q_\star)-G_\star(P_\star-Q_\star).$  Plugging \eqref{Eq:dZt-dZs-dm02} into \eqref{Eq:Gstar} we obtain the following explicitly finite propagator equation:
\beq
&&G_\star^{-1}(K_\star)=\omega_\star^2+\k^2+m_\star^2 \nonumber \\
&&\qquad - \frac{(N+2)\lambda_\star^2}{18N^2} \left[
\int_{Q_\star}^{T_\star}{\cal B}_\star(Q_\star)\delta G_\star(Q_\star;K_\star,\tilde K_\star) \right. \nonumber\\
&&\qquad\qquad+\frac{\omega_\star^2-\tilde\omega_\star^2}{\hat\omega^2_\star-\tilde\omega^2_\star}\int_{Q_\star}^{T_\star}{\cal B}_\star(Q_\star) \delta G_\star(Q_\star;\tilde K_\star,\hat K_\star)
\nonumber\\
&&\nonumber\\
&&\left.\qquad\qquad+ \frac{\k^2-\tilde\k^2}{\hat\k^2-\tilde\k^2}\int_{Q_\star}^{T_\star}{\cal B}_\star(Q_\star) \delta G_\star(Q_\star;\tilde K_\star,\bar K_\star)\right].\qquad
\label{Eq:Gstar_finite}
\eeq
The above self-consistent integral equation for $G_\star$ is solved iteratively and its solution is used in \eqref{Eq:dZt-dZs-dm02} to obtain the value of $\delta Z_\tau,$ $\delta Z_s,$ and $\delta m_0^2$.\\

\subsection{Determination of $\lambda_0^{(A/B)}$ and $\lambda_4$}
To determine $\lambda_0^{(A)}$ and $\lambda_0^{(B)}$ we impose the consistency conditions 
\beq
\bar V_\star^B(R_\star,S_\star) = \frac{\lambda_\star}{3N},\quad
\bar V_\star^S(R_\star,S_\star) = \frac{N+2}{3N}\lambda_\star,\quad
\label{Eq:ren_cond_VAB}
\eeq
and for simplicity we choose $R_\star=S_\star=0.$ Then from \eqref{Eq:bVB} and \eqref{Eq:bLB} we obtain
\begin{widetext}
\beq
\lambda_0^{(B)}=\left[\lambda_\star+\frac{\lambda_\star^2(N+6)}{6N} \left({\cal B}_\star[G_\star]-\int_{Q_\star}^{T_\star} {\cal B}_\star[G_\star](Q_\star)G_\star^2(Q_\star) \bar V^B_\star(Q_\star,0)\right) \right]
\left[1-\int_{Q_\star}^{T_\star} G_\star^2(Q_\star) \bar V_\star^B(Q_\star,0)\right]^{-1},\qquad
\label{Eq:l0B_at_zero_mom}
\eeq
with ${\cal B}_\star[G_\star]\equiv {\cal B}_\star(Q_\star=0),$ while from \eqref{Eq:bVS} and \eqref{Eq:bLS} we obtain
\beq
\frac{N\lambda_0^{(A)}+\lambda_0^{(B)}}{(N+2)\lambda_\star}=\left[1+\frac{\lambda_\star}{N}\left({\cal B}_\star[G_\star]-\frac{1}{2}\int_{Q_\star}^{T_\star} {\cal B}_\star[G_\star](Q_\star)G_\star^2(Q_\star) \bar V^S_\star(Q_\star,0)\right)\right]
\left[1-\frac{1}{2}\int_{Q_\star}^{T_\star} G_\star^2(Q_\star) \bar V_\star^S(Q_\star,0)\right]^{-1}.\qquad\quad
\label{Eq:l0A_at_zero_mom}
\eeq
\end{widetext}
Therefore, in order to have $\lambda_0^{(A/B)}$ we need to determine first $\bar V_\star^B(Q_\star,0)$ and $\bar V_\star^S(Q_\star,0),$ which is obtained by using the conditions \eqref{Eq:ren_cond_VAB} with $R_\star=S_\star=0$ in \eqref{Eq:bVB} and \eqref{Eq:bVS} taken at $T_\star,$ $\phi=0$ and one vanishing momentum. We obtain
\beq
\bar V_\star^B(Q_\star,0)&=&\frac{\lambda_\star}{3N}-\frac{\lambda_\star^2(N+6)}{18N^2}\big[{\cal B}_\star(Q_\star)-{\cal B}_\star[G_\star]\big]\nonumber\\
&-&\frac{\lambda_\star^2(N+6)}{18N^2}\left[ \int_{P_\star}^{T_\star} {\cal B}_\star(P_\star) G_\star^2(P_\star) \bar V^B_\star(P_\star,0) \right.\nonumber\\
&-&\left.\int_{P_\star}^{T_\star} {\cal B}_\star(Q_\star-P_\star) G_\star^2(P_\star) \bar V_\star^B(P_\star,0) \right],\ \ \ 
\label{Eq.VsB_final}
\eeq
and
\beq
\bar V_\star^S(Q_\star,0)&=&\frac{N+2}{3N}\lambda_\star-\frac{\lambda_\star^2(N+2)}{3N^2}\big[{\cal B}_\star(Q_\star)-{\cal B}_\star[G_\star]\big]\nonumber\\
&-&\frac{\lambda_\star^2(N+2)}{6N^2}\left[\int_{P_\star}^{T_\star} {\cal B}_\star(P_\star) G_\star^2(P_\star) \bar V^S_\star(P_\star,0)\right.\nonumber\\
&-&\left.\int_{P_\star}^{T_\star} {\cal B}_\star(Q_\star-P_\star) G_\star^2(P_\star) \bar V_\star^S(P_\star,0) 
\right],
\label{Eq.VsS_final}
\eeq
where the first integral is a convolution, while the second one is a volume-type integral. These integrals are computed numerically according to the formulas in Eqs.(114) and (115) of \cite{Marko:2012wc}. The self-consistent equations \eqref{Eq.VsB_final} and \eqref{Eq.VsS_final} can be solved iteratively using the solution of \eqref{Eq:Gstar_finite} and with their solution we can compute the volume-type integrals in \eqref{Eq:l0B_at_zero_mom} and \eqref{Eq:l0A_at_zero_mom} to determine $\lambda_0^{(B)}$ and $\lambda_0^{(A)}.$\\

To determine $\lambda_4$, we use \eqref{Eq:hatV_spec_trunc2} and impose the renormalization condition
\beq
\hat V_\star =\frac{\lambda_\star}{3N}.
\eeq
Since $\bar V_\star^A(0;0)=\bar V_\star^B(0;0)=\lambda_\star/(3N)$ and $\bar\Lambda_\star^A(0;0)$ and $\bar\Lambda_\star^B(0;0)$ can be read off from \eqref{Eq:bLA} and \eqref{Eq:bLB}, respectively, one ends up with
\beq
\lambda_4=\lambda_0^{(A)}+2\lambda_0^{(B)}-2\lambda_\star-\frac{\lambda_\star^2}{3N}(N+8){\cal B}_\star[G_\star].\qquad
\label{Eq:l4_explicit}
\eeq
Note that one can compute ${\cal B}_\star[G_\star]$ as a volume-type integral or by taking the result of a convolution at the first available value of the absolute value of the 3-momenta.

\section{Finite localized approximations}\label{app:loc}
In this Appendix, after considering two specific examples, we show how to renormalize the $\smash{N=1}$ localized bare gap equation to all orders using temperature and field-independent counterterms. The resulting recipe is very simple and amounts to replacing bare parameters by renormalized ones and local divergent integrals ${\cal D}$ by corresponding finite integrals ${\cal D}_{\rm F}$ related to ${\cal D}$ through a systematic procedure. We also explain why our renormalization procedure does not work at $N\neq 1$ or when coupling the gap equation to the field equation, and how it is nevertheless possible to define finite localized equations in these cases by extending the recipe obtained in the case of the gap equation at $N=1$.

\subsection{$N=1$: finite localized two-loop gap equation}
The localized bare two-loop gap equation in the case $N=1$ reads (we set $\bar M^2\equiv\bar M^2_{\rm L}$)
\beq
\bar M^2=m_0^2+\frac{\lambda_0}{2}{\cal T}[\bar G]+\frac{\phi^2}{2}\left(\lambda_2-\lambda_3^2{\cal B}[\bar G]\right),
\label{eq:2LN1LocGapBare}
\eeq
where we have introduced a bare coupling $\lambda_3$ for the bubble diagram, which will be needed later. This equation contains both quadratic and logarithmic divergences. We eliminate the former by reparametrizing the equation in terms of the renormalized mass at a fixed value of the field, $\phi=0$ for simplicity, and a fixed temperature $T_\star$:
\beq
m^2_\star\equiv \bar M^2_{\phi=0,T_\star}=m_0^2+\frac{\lambda_0}{2}{\cal T}_\star[G_\star]\,,\label{eq:c0}
\eeq
where $G_\star(Q_\star)\equiv 1/(Q^2_\star+m^2_\star)$. Plugging the expression for $m^2_0$ derived from (\ref{eq:c0}) back into \eqref{eq:2LN1LocGapBare}, we arrive at
\beq
\bar M^2=m_\star^2+\frac{\lambda_0}{2}\left({\cal T}[\bar G]-{\cal T}_\star[G_\star]\right)+\frac{\phi^2}{2}\left(\lambda_2-\lambda_3^2{\cal B}[\bar G]\right).
\nonumber\\\label{eq:Gap_toto}
\eeq
The contributions ${\cal T}[\bar G]-{\cal T}_\star[G_\star]$ and ${\cal B}[\bar G]$ still contain logarithmic divergences. Those in ${\cal B}[\bar G]$ will be treated in a moment. Let us first isolate and eliminate those in ${\cal T}[\bar G]-{\cal T}_\star[G_\star]$. To this purpose, we note first that the logarithmic divergences in this difference originate from the subleading contribution to $\bar G$ in an expansion around $G_\star$: $\bar G-G_\star=(\bar M^2-m^2_\star)\partial G_\star/\partial m^2_\star+\dots$. Second, once this subleading contribution to $\bar G$ has been extracted, we can replace the temperature in the outer integral ${\cal T}[\bar G]$ by $T_\star$, without affecting the divergence. It follows that the logarithmic divergence in ${\cal T}[\bar G]-{\cal T}_\star[G_\star]$ is equal to that in $(\bar M^2-m^2_\star)\partial {\cal T}_\star[G_\star]/\partial m^2_\star$. Subtracting this contribution (multiplied by $\lambda_0/2$) from both sides of (\ref{eq:Gap_toto}) and after some simple algebra, we arrive at
\beq
\bar M^2 & = & m^2_\star+\frac{\lambda_0}{2d_\star[G_\star]}{\cal T}_{\rm F}[\bar G]+\frac{\phi^2}{2d_\star[G_\star]}\Big(\lambda_2-\lambda^2_3{\cal B}[\bar G]\Big),\label{eq:d0}\nonumber\\
\eeq
where we have introduced
\beq
d_\star[G_\star]\equiv 1-\frac{\lambda_0}{2}\frac{d{\cal T}_\star[G_\star]}{dm^2_\star}\,,
\eeq
as well as the finite tadpole integral given in \eqref{eq:TadF}. The combination of integrals in \eqref{eq:TadF} is finite, as we have argued above or as can be checked by a direct calculation. To make the corresponding contribution in (\ref{eq:d0}) finite we impose that
\beq
\frac{\lambda_0}{d_\star[G_\star]}=\lambda_\star\,.
\eeq
In this way we have eliminated the logarithmic divergences present originally in ${\cal T}[\bar G]-{\cal T}_\star[G_\star]$. To get rid of the remaining divergences present in ${\cal B}[\bar G]$, we choose $\lambda_2$ and $\lambda_3$ such that
\beq
\frac{\lambda_3^2}{d_\star[G_\star]} & = & \lambda^2_\star\,,\\
\frac{\lambda_2}{d_\star[G_\star]} & = & \lambda_\star+\lambda^2_\star{\cal B}_\star[G_\star]\,,
\eeq
and the renormalized localized equation reads finally
\beq
\bar M^2=m_\star^2+\frac{\lambda_\star}{2}{\cal T}_{\rm F}[\bar G]+\frac{\phi^2}{2}\left(\lambda_\star-\lambda_\star^2{\cal B}_{\rm F}[\bar G]\right),
\label{eq:2LN1LocGapFin}
\eeq
where the finite bubble integral is defined in \eqref{eq:BubF}.

When comparing the localized bare gap equation (\ref{eq:2LN1LocGapBare}) and its renormalized form (\ref{eq:2LN1LocGapFin}), we observe that the rule to pass from one form to the other is to replace bare parameters by renormalized ones and each local diagram ${\cal D}$ by a finite version ${\cal D}_{\rm F}$.

\subsection{$N=1$: finite localized ${\cal O}(\lambda^2)$ gap equation}\label{app:N1BBGap}
The same strategy can be used to renormalize the bare localized ${\cal O}(\lambda^2)$ gap equation. In this case the localized bare gap equation is
\beq
\bar M^2=m_0^2+\frac{\lambda_0}{2}{\cal T}[\bar G]+\frac{\phi^2}{2}\left(\lambda_2-\lambda_3^2{\cal B}[\bar G]\right)-\frac{\lambda_4^2}{6}{\cal S}[\bar G]\,.\nonumber\\
\label{eq:Ol2N1LocGapBare2}
\eeq
To absorb the quadratic divergences, we set
\beq
m_0^2=m_\star^2-\frac{\lambda_0}{2}{\cal T}_\star[G_\star]+\frac{\lambda_4}{6}{\cal S}_\star[G_\star]\,.
\eeq
After the expression for $m^2_0$ is used in (\ref{eq:Ol2N1LocGapBare2}), there remain logarithmic divergences in ${\cal B}[\bar G]$, ${\cal T}[\bar G]-{\cal T}_\star[G_\star]$ and ${\cal S}[\bar G]-{\cal S}_\star[G_\star]$. We need to distinguish two types of such divergences. Those which originate from four-point subgraphs drawn on the original diagrams in terms of the self-consistent propagator $\bar G$, and those which originate from four-point subgraphs drawn on the diagrams obtained after iterating the gap equation a certain number of times. The logarithmic divergence in ${\cal B}[\bar G]$ is of the first type, while the logarithmic divergence in ${\cal T}[\bar G]-{\cal T}_\star[G_\star]$ is of the second type and ${\cal S}[\bar G]-{\cal S}_\star[G_\star]$ contains both types of divergences. The second type of logarithmic divergences, which we shall deal with first, has to do with the subleading term in the expansion of $\bar G$ around $G_\star$. More precisely, for a given diagram ${\cal D}[\bar G]$, a reasoning similar to the one above shows that the divergence of the second type contained in ${\cal D}[\bar G]$ takes the form $(\bar M^2-m^2_\star)d{\cal D_\star}[G_\star]/dm^2_\star$. We then subtract these type of contributions, for ${\cal D}={\cal T}$ and ${\cal D}={\cal S}$ (multiplied with the corresponding coupling factors) from both sides of the gap equation. After some simple algebra, we arrive at
\beq\label{eq:e0}
\bar M^2 & = & m_\star^2+\frac{\phi^2}{2}\left(\frac{\lambda_2}{d_\star[G_\star]}-\frac{\lambda_3^2}{d_\star[G_\star]}{\cal B}[\bar G]\right)\nonumber\\
& + & \frac{1}{2}\frac{\lambda_0}{d_\star[G_\star]}\left({\cal T}[\bar G]-{\cal T}_\star[G_\star]-(\bar M^2-m^2_\star)\frac{d{\cal T}_\star}{dm^2_\star}\right)\nonumber\\
& - & \frac{1}{6}\frac{\lambda_4^2}{d_\star[G_\star]}\left({\cal S}[\bar G]-{\cal S}_\star[G_\star]-(\bar M^2-m^2_\star)\frac{d{\cal S}_\star}{dm^2_\star}\right),\nonumber\\
\eeq
where we have introduced
\beq
d_\star[G_\star]\equiv 1-\frac{\lambda_0}{2}\frac{d{\cal T}_\star[G_\star]}{dm_\star^2}+\frac{\lambda_4^2}{6}\frac{d{\cal S}_\star[G_\star]}{dm_\star^2}.
\label{eq:d}
\eeq
We need finally to deal with the logarithmic divergences of the first type. Those in the first line of (\ref{eq:e0}) are similar to the ones discussed in the two-loop case. They are absorbed by choosing $\lambda_2$ and $\lambda_3$ such that
\beq
\label{Eq:l3wd}
\frac{\lambda_3^2}{d_\star[G_\star]} & = & \lambda^2_\star\,,\\
\frac{\lambda_2}{d_\star[G_\star]} & = & \lambda_\star+\lambda^2_\star{\cal B}_\star[G_\star]\,.
\eeq
The round bracket in the second line of (\ref{eq:e0}) does not contain more divergences. The third line does and it is easy to convince oneself that they are absorbed by choosing $\lambda_0$ and $\lambda_4$ such that
\beq
\frac{\lambda_4^2}{d_\star[G_\star]} & = & \lambda^2_\star\,,\\
\frac{\lambda_0}{d_\star[G_\star]} & = & \lambda_\star+\lambda^2_\star{\cal B}_\star[G_\star]\,.
\label{Eq:l0wd}
\eeq
All in all, we arrive at the following finite equation
\beq\label{eq:2LN1LocGapFin2}
\bar M^2=m_\star^2+\frac{\lambda_\star}{2}{\cal T}_{\rm F}[\bar G]+\frac{\phi^2}{2}\left(\lambda_\star-\lambda_\star^2{\cal B}_{\rm F}[\bar G]\right)-\frac{\lambda_\star^2}{6}{\cal S}_{\rm F}[\bar G]\,,\nonumber\\
\eeq
where ${\cal T}_{\rm F}[\bar G],$ ${\cal B}_{\rm F}[\bar G]$ and ${\cal S}_{\rm F}[\bar G]$  were defined respectively in \eqref{eq:TadF}, \eqref{eq:BubF}, and \eqref{eq:SSF}.

Again, when comparing the localized bare gap equation (\ref{eq:Ol2N1LocGapBare2}) and its renormalized form (\ref{eq:2LN1LocGapFin2}), we observe that the rule to pass from one form to the other is to replace bare parameters by renormalized ones and each local diagram ${\cal D}$ by a finite version ${\cal D}_{\rm F}$.

We note that $\smash{\lambda_2=\lambda_0}$ and $\smash{\lambda_4=\lambda_3}$. This comes as no surprise, as we discuss in the next subsection. We mention finally that, because $d_\star[G_\star]$ depends on $\lambda_0$ and $\lambda_4$, the above expressions give the bare parameters implicitly. We can obtain explicit expressions in the following way. We write
\beq
d_\star[G_\star]=\frac{\lambda_0}{\lambda_\star+\lambda^2_\star{\cal B}_\star[G_\star]}=\frac{\lambda^2_4}{\lambda^2_\star}\,,
\eeq
and note that, if $a/b=c/d$, we have
\beq
\frac{a}{b}=\frac{\alpha a+\beta c}{\alpha b+\beta d}
\eeq
for any $\alpha$ and $\beta$ different from zero. Choosing $\alpha=(-1/2)d{\cal T}_\star[G_\star]/dm^2_\star$ and $\beta=(1/6)d{\cal S}_\star[G_\star]/dm^2_\star$, we arrive at
\beq
d_\star[G_\star]=\frac{d_\star[G_\star]-1}{-\frac{1}{2}\left(\lambda_\star+\lambda^2_\star{\cal B}_\star[G_\star]\right)\frac{d{\cal T}_\star[G_\star]}{dm^2_\star}+\frac{\lambda^2_\star}{6}\frac{d{\cal S}_\star[G_\star]}{dm^2_\star}}\nonumber\\
\eeq
or
\beq
d_\star[G_\star]b_\star[G_\star]=1
\label{Eq:db1}
\eeq
with
\beq
b_\star[G_\star]\equiv1+\frac{1}{2}(\lambda_\star+\lambda_\star^2{\cal B}_\star[G_\star])\frac{d{\cal T}_\star[
G_\star]}{dm^2_\star}-\frac{\lambda_\star^2}{6}\frac{d{\cal S}_\star[G_\star]}{dm_\star^2}\,.\nonumber\\
\eeq
Using \eqref{Eq:db1} in \eqref{Eq:l3wd}-\eqref{Eq:l0wd}, it follows immediately that
\beq
\lambda_0&=&\lambda_2=\frac{\lambda_\star+\lambda_\star^2{\cal B}_\star[G_\star]}{b_\star[G_\star]}\,,\\
\lambda_3^2&=&\lambda_4^2=\frac{\lambda_\star^2}{b_\star[G_\star]}\,.
\eeq

\subsection{$N=1$ gap equation: all-order renormalization \label{ss:renall}}
Let us now see how to generalize the above procedure to all orders. Let us write the gap equation in a certain truncation in the form
\beq
\bar M^2&=&m_0^2+\sum_{{\cal D}_0}\lambda_{{\cal D}_0}^{n_{{\cal D}_0}}\,{\cal D}_0[\bar G]+\sum_{{\cal D}}\lambda_{{\cal D}}^{n_{{\cal D}}}\,{\cal D}[\phi,\bar G]\,,
\label{eq:ProofGapBare}
\eeq
where ${\cal D}_0$ runs over the diagrams containing no explicit dependence on the field and ${\cal D}$ runs over the remaining diagrams. 
Power counting tells us that ${\cal D}_0$-type diagrams contain quadratic divergences. To get rid of those we rewrite the gap equation in terms of the renormalized mass at a given value of the field, $\phi=0$ for simplicity, and a given value of the temperature, $T_\star$:
\beq
\bar M^2_{\phi=0}(T=T_\star)\equiv m_\star^2=m_0^2+\sum_{{\cal D}_0}\lambda_{{\cal D}_0}^{n_{{\cal D}_0}}\,{\cal D}_{0,\star}[G_\star]\,.\nonumber\\
\label{eq:Proof_m02}
\eeq
After this reparametrization, there are still logarithmic divergences in the gap equation, both in differences ${\cal D}_0[\bar G]-{\cal D}_{0,\star}[G_\star]$ involving field-independent diagrams and in the field-dependent ones. Those divergences are of two types: divergences associated to four-point graphs that one can draw on the original diagrams in terms of self-consistent propagators $\bar G$, and divergences originating from the subleading contribution to $\bar G$ in an expansion around $G_\star$.\footnote{These divergences are associated to the four-point graphs that one can draw on the diagrams obtained after iterating the gap equation a certain number of times} The field-dependent diagrams contain only logarithmic divergences of the first type, whereas the differences ${\cal D}_0[\bar G]-{\cal D}_{0,\star}[G_\star]$ contain both logarithmic divergences of the first and of the second type.\footnote{Except for the case of the tadpole diagram (${\cal D}_0={\cal T}$) where the divergences are only of the second type.} 

Following the same line of thought as in the examples above, it is simple to convince oneself that the logarithmic divergence of the second type contained in ${\cal D}_0[\bar G]-{\cal D}_{0,\star}[G_\star]$, is nothing but $(\bar M^2-m^2_\star)\partial{\cal D}_{0,\star}[G_\star]/\partial m^2_\star$ and that it can be eliminated by subtracting it from both sides of the gap equation. After some simple algebra, one arrives at
\begin{widetext}
\beq\label{eq:first}
\bar M^2&=&m_\star^2+\sum_{{\cal D}_0}\frac{\lambda_{{\cal D}_0}^{n_{{\cal D}_0}}}{d_\star[G_\star]}\,\left({\cal D}_0[\bar G]-{\cal D}_{0,\star}[G_\star]-(\bar M^2-m^2_\star)\frac{\partial{\cal D}_{0,\star}[G_\star]}{\partial m^2_\star}\right)+\sum_{{\cal D}}\frac{\lambda_{{\cal D}}^{n_{{\cal D}}}}{d_\star[G_\star]}\,{\cal D}[\phi,\bar G]\,,
\eeq
\end{widetext}
where we have introduced
\beq
d_\star[G_\star]\equiv 1-\sum_{{\cal D}_0}\lambda_{{\cal D}_0}^{n_{{\cal D}_0}}\frac{\partial {\cal D}_{0,\star}[G_\star]}{\partial m_\star^2}\,,
\label{eq:d_redef}
\eeq
generalizing the function introduced in the examples above (we employ the same notation here as it is obvious that we are dealing with a different approximation and thus a different function). 

It remains to get rid of the logarithmic divergences of the first type that are still present in (\ref{eq:first}). Let us start dealing with those present in the field-dependent diagrams ${\cal D}[\phi,\bar G]$. To this purpose, one applies the standard Bogoliubov-Parasiuk-Hepp-Zimmermann (BPHZ) procedure. To each diagram $\lambda_\star^{n_{\cal D}}{\cal D}[\phi,\bar G]$, the BPHZ procedure associates a family of lower loop diagrams $\delta\hat\lambda_{{\cal D}',{\cal D}}\,{\cal D}'[\phi,\bar G]$ whose role is to remove consistently all the divergences of the original diagram. Each of these lower diagrams is multiplied by a ``counterterm'' factor $\delta\hat\lambda_{{\cal D}',{\cal D}}$ depending on both ${\cal D}'$ and ${\cal D}$ and determined in a specific way according to the BPHZ procedure.\footnote{To be consistent with the determination of the other bare parameters, we choose a scheme where the counterterms are evaluated at zero momentum, zero field and temperature $T_\star$, in a way such that, in particular, the propagator that enters the counterterms is $G_\star$.} The $\Phi$-derivable truncations that we consider in general are such that, given the higher loop diagrams present in the truncation, the lower loop diagrams associated through the BPHZ procedure are also present in the truncation. Then, in order to renormalize $\sum_{{\cal D}} (\lambda^{n_{\cal D}}/d_\star[G_\star]){\cal D}[\phi,\bar G]$, we need only to adjust $\lambda^{n_{\cal D}}/d_\star[G_\star]$ to:
\beq\label{eq:ct1}
\frac{\lambda^{n_{\cal D}}}{d_\star[G_\star]}=\lambda_\star^{n_{\cal D}}+\delta\hat\lambda_{{\cal D}}\,,
\eeq
with
\beq
\delta\hat\lambda_{{\cal D}}=\sum_{{\cal D}'}\delta\hat\lambda_{{\cal D},{\cal D}'}\,,
\eeq
where the sum runs over the diagrams ${\cal D}'$ present in the truncation and whose BPHZ renormalization involves the diagram ${\cal D}$. Obviously, for the higher loop diagrams of the truncation, $\delta\hat\lambda_{\cal D}=0$ because these diagrams do not absorb any divergence. With not much effort, it is possible to argue that the choice
\beq\label{eq:ct2}
\frac{\lambda^{n_{{\cal D}_0}}}{d_\star[G_\star]}=\lambda_\star^{n_{{\cal D}_0}}+\delta\hat\lambda_{{\cal D}_0}
\eeq
with
\beq
\delta\hat\lambda_{{\cal D}_0}=\sum_{{\cal D}'_0}\delta\hat\lambda_{{\cal D}_0,{\cal D}'_0}\,,
\eeq
and where the sum runs over all the diagrams appearing in the BPHZ renormalization of the explicit (first type) four-point divergences of ${\cal D}_0$ and $\delta\hat\lambda_{{\cal D}'_0,{\cal D}_0}$, removes the remaining logarithmic divergences of the first type in the field-independent part of Eq.~(\ref{eq:first}). In fact, after some simple rewriting, we arrive at 
\beq
\bar M^2&=&m_\star^2+\sum_{{\cal D}_0}\lambda^{n_{{\cal D}_0}}_\star{\cal D}_{0,F}[\bar G]+\sum_{{\cal D}}\lambda^{n_{\cal D}}_\star{\cal D}_{\rm F}[\phi,\bar G]\,,\nonumber\\
\eeq
with
\begin{widetext}
\beq
{\cal D}_{0,\rm F}[\bar G] & = & \left({\cal D}_0[\bar G]-{\cal D}_{0,\star}[G_\star]-(\bar M^2-m^2_\star)\frac{\partial{\cal D}_{0,\star}[G_\star]}{\partial m^2_\star}\right)+\sum_{{\cal D}'_0}\frac{\delta\hat\lambda_{{\cal D}'_0,{\cal D}_0}}{\lambda^{n_{{\cal D}_0}}_\star}\left({\cal D}'_0[\bar G]-{\cal D}'_{0,\star}[G_\star]-(\bar M^2-m^2_\star)\frac{\partial{\cal D}'_{0,\star}[G_\star]}{\partial m^2_\star}\right)\nonumber\\\\
{\cal D}_{\rm F}[\phi,\bar G] &=& {\cal D}[\phi,\bar G]+\sum_{{\cal D}'}\frac{\delta\hat\lambda_{{\cal D}',{\cal D}}}{\lambda^{n_{\cal D}}_\star}{\cal D}'[\phi,\bar G]\,,
\eeq
\end{widetext}
where the sums $\sum_{{\cal D}'_0}$ and $\sum_{{\cal D}'}$ run over the diagrams that appear in the BPHZ renormalization of the explicit (first type) four-point divergences contained in ${\cal D}_0[\bar G]$ and ${\cal D}[\phi,\bar G]$, making ${\cal D}_{0,\rm F}[\bar G]$ and ${\cal D}_{\rm F}[\phi,\bar G]$ the natural finite versions of ${\cal D}_0[\bar G]$ and ${\cal D}[\phi,\bar G]$. 
 
We note that the above expressions (\ref{eq:ct1}) and (\ref{eq:ct2}) only define the bare parameters implicitly. If we want to obtain explicit expressions, we can proceed as above, writing
\beq
d_\star[G_\star]=\frac{\lambda^{n_{{\cal D}_0}}}{\lambda_\star^{n_{{\cal D}_0}}+\delta\hat\lambda_{{\cal D}_0}}\,,
\eeq
for any ${\cal D}_0$. Then, owing to the fact that
\beq
u=\frac{a_i}{b_i} \quad \forall i
\eeq
implies that
\beq
u=\frac{\sum_i\alpha_i a_i}{\sum_i \alpha_i b_i}\,,
\eeq
we conclude that
\beq
d_\star[G_\star]=\frac{1-d_\star[G_\star]}{\sum_{{\cal D}_0}(\lambda_\star^{n_{{\cal D}_0}}+\delta\hat\lambda_{{\cal D}_0})\frac{d{\cal D}_{0,\star}[G_\star]}{dm^2_\star}}
\eeq
that is
\beq
d_\star[G_\star]b_\star[G_\star]=1\,,
\eeq
with
\beq
b_\star[G_\star]=1+\sum_{{\cal D}_0}(\lambda_\star^{n_{{\cal D}_0}}+\delta\hat\lambda_{{\cal D}_0})\frac{\partial {\cal D}_{0,\star}[G_\star]}{\partial m_\star^2}\,.
\eeq
The relations (\ref{eq:ct1}) and (\ref{eq:ct2}) are then simply inverted and we obtain
\begin{subequations}
\beq
\lambda_{{\cal D}_0}^{n_{{\cal D}_0}}&=&\frac{\lambda_\star^{n_{{\cal D}_0}}+\delta\hat\lambda_{{\cal D}_0}}{b_\star[G_\star]}\,,\\
\lambda_{{\cal D}}^{n_{{\cal D}}}&=&\frac{\lambda_\star^{n_{{\cal D}}}+\delta\hat\lambda_{{\cal D}}}{b_\star[G_\star]}\,.
\eeq
\end{subequations}
We note finally that, if the truncation is such that for each diagram ${\cal D}_0[G]$ present in truncation the diagram ${\cal D}'[\phi,G]=\phi^2\delta {\cal D}_0/\delta G(0)$ is also present in the truncation, and vice versa, then the corresponding bare couplings coincide: $\lambda_{{\cal D}_0}=\lambda_{{\cal D}'}$. This is for instance the case of the localized ${\cal O}(\lambda^2)$ truncation discussed in Appendix~\ref{app:N1BBGap} and explains why we obtained $\lambda_4=\lambda_0$ and $\lambda_3=\lambda_2$ in this case.

\subsection{$N=1$ field equation \label{ss:fieldEqN1}}
We turn now our attention to the renormalization of the field equation. We define the localized field equation in line with the definition of a localized gap equation, that is we take the full, bare 2PI field equation and compute the diagrams with the local {\it ansatz} used in the gap equations. Note that another way to define the field equation is to take the bare 2PI effective potential with local {\it ansatz}, and differentiate it with respect to $\phi$. This would bring additional terms which originate from $\delta \gamma/\delta G|_{G=\bar G_{\rm localized}}\neq0$ as in a general truncation the localized {\it ansatz} does not satisfy the stationarity condition for the propagator. An important difference between the two approaches is that the solution of the latter is an extremum of the potential evaluated with the local {\it ansatz}, while the solution of the former is not.\footnote{This raises the question of the proper definition of the potential in the localized approach, which was discussed in  Sec.~\ref{sec:IR-sensitivity}.}\\

In the full 2PI approach the gap and field equations share the bare couplings of the effective action they both are derived from. In the localized version we lift this constraint and try to renormalize the gap and field equations by using in each equation a different set of counterterms. However, we will show that even exploiting this freedom is not enough to obtain finite equations. We illustrate this in the two-loop case, where the bare localized field equation in the broken phase (at $h=0$) reads
\beq
0=m_2^2+\frac{\tilde\lambda_4}{6}\bar\phi^2+\frac{\tilde\lambda_2}{2}{\cal T}[\bar G]-\frac{\tilde\lambda_3^2}{6}{\cal S}[\bar G]\,.
\label{eq:BareLocFieldN1}
\eeq
Following the logic of both the full 2PI treatment and the renormalization of the gap equation in the localized case, we write
\beq
m_2^2=m_\star^2-\frac{\tilde\lambda_2}{2}{\cal T}_\star[G_\star]+\frac{\tilde\lambda_3^2}{6}{\cal S}_\star[G_\star]\,.
\eeq
Using this expression in \eqref{eq:BareLocFieldN1}, in which we add $\bar M^2$ to both sides of the equation, we express the resulting difference of tadpoles and that of setting sums with the help of \eqref{eq:TadF} and \eqref{eq:SSF} to obtain
\beq
&&(\bar M^2-m_\star^2)\left(1-\frac{\tilde\lambda_2}{2}\frac{d {\cal T}_\star[G_\star]}{d m^2_\star}+\frac{\tilde\lambda_3^2}{6}\frac{d{\cal S}_\star[G_\star]}{d m_\star^2}\right)\nonumber\\
&&=\bar M^2+\frac{\tilde\lambda_4}{6}\bar\phi^2+\frac{{\cal T}_{\rm F}[\bar G]}{2}\left(\tilde\lambda_2-\tilde\lambda_3^2{\cal B}_\star[G_\star]\right)-\frac{\tilde\lambda_3^2}{6}{\cal S}_{\rm F}[\bar G]\,.\nonumber\\
\label{eq:processedBareLocFieldN1}
\eeq
Comparing \eqref{eq:BareLocFieldN1} and \eqref{eq:processedBareLocFieldN1} we see that in order to obtain a finite field equation in which the bare couplings are replaced by renormalized ones and the integrals are replaced by their UV finite form the three bare couplings $\tilde\lambda_2,$ $\tilde\lambda_3,$ and $\tilde\lambda_4$ have to obey four independent constraints which leads to contradiction. Therefore the field equation is not renormalazible with the procedure used.

The problem can also be understood in a diagrammatic way. The gap equation may contain diagrams, which are not present in the field equation. For example the two-loop gap equation \eqref{eq:2LN1LocGapBare} contains the bubble diagram, while the field equation \eqref{eq:BareLocFieldN1} does not. In the full 2PI treatment, expanding the full propagator in the diagrams of the field equation leads to diagrams with self-energy insertions. The renormalization of these diagrams is discussed in \cite{Berges:2005hc}. However in the localized version, instead of self-energy insertions, one obtains \{self-energy\} $\times$ \{original\,diagram\} type of contributions. The renormalization of these types of divergences would require \{counterterm\} $\times$ \{self-energy\} type of contributions. If there are certain diagrams in the self-energy which are not present in the field equation, these divergences cannot be absorbed by counterterms.

Even though the analysis presented here shows that counterterms cannot be chosen in such a way that they make the localized bare field equation finite, we can always define a finite localized field equation by replacing all the couplings by renormalzied one an all integrals by finite ones, following the rule discussed in Appendix~\ref{ss:renall}.\\

\subsection{$N\neq1$ gap equations}

By examining the coupled gap equations at $N\neq1$ we find a similar obstruction to the one in the $N=1$ field equation. We illustrate this using the two-loop truncation, where the coupled, localized bare gap equations read
\beq
\bml & = & m^2_0+\frac{\lambda^{(A+2B)}_0}{6N}{\cal T}[\bgl]+\frac{\lambda^{((N-1)A)}_0}{6N}{\cal T}[\bgt]\nonumber\\
&&+\frac{\phi^2}{6N}\big[\lambda_2^{(A+2B)}-\frac{\lambda_3^2}{3N}\big(9{\cal B}[\bgl]+(N-1){\cal B}[\bgt]\big)\big]\,,\nonumber\\
\label{eq:bml_loc2L}\\
\bmt & = & m^2_0+\frac{\lambda^{(A)}_0}{6N}{\cal T}[\bgl]+\frac{\lambda^{((N-1)A+2B)}_0}{6N}{\cal T}[\bgt]\nonumber\\
&&+\frac{\phi^2}{6N}\big(\lambda_2^{(A)}-\frac{2\tilde\lambda_3^2}{3N}{\cal B}[\bgl;\bgt]\big)\,.
\label{eq:bmt_loc2L}
\eeq
Following the steps of the $N=1$ case, we remove the quadratic divergences by choosing
\beq
m_0^2=m_\star^2-\frac{\lambda_0^{(NA+2B)}}{6N}{\cal T}_\star[G_\star]\,.
\eeq
It is then convenient to consider the combinations $\bml+(N-1)\bmt$ and $\bml-\bmt$. In the equations for these combinations only certain combinations of bare couplings appear, $\lambda_{0,2}^{(NA+2B)}$ and $\lambda_{0,2}^{(2B)}$ respectively (the couplings for the bubbles do not combine of course). It is then pretty clear that the logarithmic divergences hidden in the tadpoles can be absorbed by setting
\beq
\frac{1}{\lambda_0^{(NA+2B)}} &=&\frac{1}{(N+2)\lambda_\star}-\frac{1}{6N}{\cal B}_\star[G_\star]\,,\\
\frac{1}{\lambda_0^{(B)}}&=&\frac{1}{\lambda_\star}-\frac{1}{3N}{\cal B}_\star[G_\star]\,.
\eeq
After some simple manipulations we reach
\begin{widetext}
\beq
\bml+(N-1)\bmt&=&Nm_\star^2+\lambda_\star\frac{(N+2)}{6N}\Bigg\{\big[{\cal T}_{\rm F}[\bgl]+(N-1){\cal T}_{\rm F}[\bgt]\big]\nonumber\\
&&+\frac{\phi^2}{\lambda_{0}^{(NA+2B)}}\Big[\lambda_{\rm 2,l}^{(NA+2B)}-\frac{\lambda_3^2}{3N}\big(9{\cal B}_{\rm F}[\bgl]+(N-1){\cal B}_{\rm F}[\bgt]\big)-\frac{2(N-1)}{3N}\tilde\lambda_3^2{\cal B}_{\rm F}[\bgl;\bgt]\Big]\Bigg\}\,,\nonumber\\
\bml-\bmt&=&\frac{\lambda_\star}{3N}\Bigg\{{\cal T}_{\rm F}[\bgl]-{\cal T}_{\rm F}[\bgt]+\frac{\phi^2}{\lambda_0^{(B)}}\Big[\lambda_{\rm 2,l}^{(B)}-\frac{\lambda_3^2}{6N}\big(9{\cal B}_{\rm F}[\bgl]+(N-1){\cal B}_{\rm F}[\bgt]\big)+\frac{\tilde\lambda_3^2}{3N}{\cal B}_{\rm F}[\bgl;\bgt]\Big]\Bigg\}\,.\label{Eq:combMTML}\nonumber\\
\eeq
\end{widetext}
Here we introduced $\lambda_{\rm 2,l}^{(A/B)}$ through the relations 
\beq
\lambda_2^{(NA+2B)}&=&\lambda_{\rm 2,l}^{(NA+2B)}\nonumber\\
&+&\Big(\frac{N+8}{3N}\lambda_3^2+\frac{2(N-1)}{3N}\tilde\lambda_3^2\Big){\cal B}_\star[G_\star]\,,\nonumber\\
\lambda_2^{(B)}&=&\lambda_{\rm 2,l}^{(B)}+\Bigg(\frac{N+8}{6N}\lambda_3^2-\frac{\tilde\lambda_3^2}{3N}\Bigg){\cal B}_\star[G_\star]\,.\quad
\eeq
In order to fully renormalize the equations, it is natural to set $\lambda_{\rm 2,l}^{(NA+2B)}=\lambda_{0}^{(NA+2B)}$ and $\lambda_{\rm 2,l}^{(B)}=\lambda_{0}^{(B)}$, however in \eqref{Eq:combMTML} one obtains for $\lambda_3^2$ and $\tilde\lambda_3^2$ two conditions for each quantity, which cannot be fulfilled at the same time. This contradiction illustrates once again the impossibility to renormalize the localized bare gap equations (for $N\neq 1$) using bare parameters.

This has a similar diagrammatic explanation as in the case of the $N=1$ field equation, that is, since the two gap equations are coupled when one iterates the equations, one finds in each equation divergences multiplying diagrams which are not present in the equation before iteration. For instance, in the two-loop case, when looking at the longitudinal gap equation we find a divergence of the form ${\cal B}[\bgt]\times{\cal B}_{\rm F}[\bgl;\bgt]$ which could only be renormalized by a bare coupling multiplying ${\cal B}_{\rm F}[\bgl;\bgt]$.

Again, despite this difficulty, we can nevertheless define finite localized equations by extending in a straightforward way the prescription obtained in Appendix~\ref{ss:fieldEqN1}.

\section{The zero-momentum, two-mass setting-sun diagram in cutoff regularization}\label{app:SSMmm}

In this section we give the expression of the setting-sun integral ${\cal S}[G_1;G_2;G_2]$ written in terms of two different local (free-type) propagators $G_{1,2}(Q)=1/(Q^2+M^2_{1,2})$ with masses $M_1^2$ and $M_2^2$ and computed using a 3D cutoff regularization scheme in which the modulus of the spatial part of momentum in each propagator of the integral is cut. Previously we used this regularization in \cite{Marko:2012wc}.  We start from a formula which is the generalization of (A19) in \cite{Marko:2012wc} to the present case, obtained with the method of Ref.~\cite{Blaizot:2004bg} which uses the spectral representation of the propagator. With the Matsubara sums evaluated, one has in terms of the free spectral function
\beq
\nonumber
\rho_{1,2}(q_0,q) = \frac{1}{2\varepsilon_{1,2}(q)}\left(\delta(q_0-\varepsilon_{1,2}(q))-\delta(q_0+\varepsilon_{1,2}(q))\right),\\
\eeq
where $\varepsilon_{1,2}(q) = (q^2+M_{1,2}^2)^{1/2}$, the following expression
\begin{widetext}
\beq
\nonumber
{\cal S}[G_1;G_2;G_2] = \int_{-\infty}^\infty dq_0\,\int_{-\infty}^\infty dk_0\,\int_{-\infty}^\infty dp_0 \int_{q<\Lambda}\int_{k<\Lambda}\int_{p<\Lambda} (2\pi)^3\delta^{(3)}(\vec q+\vec k+\vec p)\rho_1(q_0,q)\rho_2(k_0,k)\rho_2(p_0,p)\\
\times\frac{n_T(q_0)n_T(p_0)-n_T(q_0)n_T(-k_0)+n_T(-p_0)n_T(-k_0)}{q_0+k_0+p_0+i\alpha},
\eeq
with $n_T(q_0)=1/(\exp(q_0/T)-1).$ Exploiting the $\delta$-functions and evaluating the remaining angular integrals yields
\beq
{\cal S}[G_1;G_2;G_2] =  \frac{1}{64\pi^4}\int_0^\Lambda dq\,\int_0^\Lambda d k\frac{q k}{\varepsilon_2(k)}\Bigg\{
\frac{1}{\varepsilon_2(q)}\Bigg[\ln\left|\frac{k^2+q^2+2kqx_\Lambda+M_1^2-(\varepsilon_2(k)+\varepsilon_2(q))^2}{k^2+q^2-2kq+M_1^2-(\varepsilon_2(k)+\varepsilon_2(q))^2}\right|\nonumber \\
+2n_T(\varepsilon_2(q))\left[1+n_T(\varepsilon_2(k))\right]\ln\left|\frac{(2kqx_\Lambda+M_1^2-2M_2^2)^2-4\varepsilon_2^2(q)\varepsilon_2^2(k)}{(2kq-M_1^2+2M_2^2)^2-4\varepsilon_2^2(q)\varepsilon_2^2(k)}\right|\Bigg]\nonumber \\
+\frac{1}{\varepsilon_1(q)}\left[1+2n_T(\varepsilon_1(q))\right]\left[1+2n_T(\varepsilon_2(k))\right] \ln\left|\frac{(2kqx_\Lambda-M_1^2)^2-4\varepsilon_1^2(q)\varepsilon_2^2(k)}{(2kq+M_1^2)^2-4\varepsilon_1^2(q)\varepsilon_2^2(k)}\right|
\Bigg\},
\label{Eq:SSMmm_3d_cut}
\eeq
where $x_\Lambda = (\Lambda^2-k^2-q^2)/2kq$ if $k \geq \Lambda-q$ and $1$ otherwise. We compute the remaining double integral numerically, using the doubly-adaptive CQUAD routine of the GNU Scientific Library (GSL) \cite{GSL}. However, to facilitate these computations, we integrate the logarithmic divergences analytically. That is, when the argument of the logarithm in an integrand of the form $f(x)\log g(x)$ vanishes at some $x_0=0$ in the integration region, we write
\beq
f(x)\ln g(x) &=& f(x)\ln g(x)-f(x_0)\ln|x-x_0|+f(x_0)\ln|x-x_0|,
\eeq
and integrate analytically the expression of the second line, while the expression of the first line, which is a smooth function around $x_0$, is integrated numerically. Also it is important to reduce the arguments of the logarithms as much as possible, because the difference of large numbers inside a logarithm may easily lead to rounding errors.
\end{widetext}


\end{document}